\documentclass[aps,prb]{revtex4}

\usepackage{epsfig}
\usepackage{amsmath}

\newcommand{\tr}{\mathop{\rm tr}}
\newcommand{\sign}{\mathop{\rm sign}}

\newcommand{\ch}{\mathop{\rm ch}}
\newcommand{\sh}{\mathop{\rm sh}}

\begin{document}

\title{Many-particle correlations in non-equilibrium Luttinger liquid}
\author{I.V. Protopopov$^{1,2}$,  D. B. Gutman$^{3}$, and A. D. Mirlin$^{1,4,5}$}
\affiliation{
\mbox{$^1$Institut f\"ur Nanotechnologie, Karlsruhe Institute of Technology, 
 76021 Karlsruhe, Germany}\\
\mbox{$^2$L.D. Landau Institute for Theoretical Physics, Kosygin
  str. 2, 119334 Moscow, Russia }\\ 
\mbox{$^3$Department of Physics, Bar Ilan University, Ramat Gan 52900,
Israel }\\
\mbox{$^4$Institut f\"ur Theorie der kondensierten Materie and DFG
  Center for Functional Nanostructures,}
\mbox{Karlsruhe Institute of Technology, 76128 Karlsruhe, Germany}\\
\mbox{$^5$Petersburg Nuclear Physics Institute, 188300 St.~Petersburg, Russia}
}
\begin{abstract}
We develop an operator-based approach to the problem of Luttinger
liquid conductor in a non-equilibrium stationary state. We show that the
coherent-state many-body fermionic density matrix as well as all fermionic
correlation functions out of equilibrium are  given by
one-dimensional functional determinants of the Fredholm type. 
Thus, the  model constitutes a remarkable example of a many-body 
problem where all the correlation functions can be evaluated exactly.
On the basis
of the general formalism we investigate four-point correlation
functions  of the fermions coming out of the Luttinger liquid
wire. Obtained correlations in the fermionic distribution functions
represent the combined effect of interaction and non-equilibrium
conditions. 
\end{abstract}
\maketitle

\section{Introduction}
It was understood long ago \cite{Luttinger} that conventional description of 
interacting fermionic systems in the framework of Landau Fermi-liquid
theory is not applicable in one dimension (1D) because of infrared
divergences. In particular, the second order of perturbation theory
yields a singularity in the self-energy at the Fermi surface, 
$\Re\Sigma\sim\epsilon\ln\epsilon$, implying a vanishing
quasi-particle weight. This violates the one-to-one correspondence
(which plays the central role in the construction of the Landau
theory) between
unperturbed electronic states and elementary excitations of the
interacting system. The lack of correspondence is a hallmark of
the emerging strongly correlated electronic state - Luttinger liquid
(LL). 

The physics of LL is known to be relevant for many systems available
in experiment. The applications of this concept include carbon
nanotubes \cite{Bockrath,Yao, Schoenenberger}, semiconducting and
metallic nanowires \cite{Auslaender, Slot}, edge states of the samples
in the quantum Hall regime \cite{Picciotto, Grayson, Chang} and spin
ladders\cite{GiamarchiTsvelik,Chitra} 
  
In view of singular infrared behavior, it is hard to access LL 
by conventional methods of many-body fermion perturbation
theory. To overcome this difficulty, a powerful approach--the
bosonization--was developed.  It is
based on the fact that in 1D fermionic creation and
annihilation operators have simple representations in terms of bosonic
fields describing such observable as charge and spin density. 
Thus, any 1D fermionic system is equivalent (as long as considered
energies are not too high)
to some (generally interacting) bosonic system. This Fermi-Bose equivalence
was first discovered on the level of correspondence between 
correlation functions in fermionic and bosonic theories  
\cite{MattisLieb,Mattis,Luther,Heidenreich,Coleman,Mandelstam}. 
Later on, underlying operator relations were derived
by Haldane \cite{Haldane}, providing the solid basis for
bosonization (for a recent detailed exposition, see \cite{Delft}). 
 
Bosonization has proven to be a very efficient tool for tackling 1D
interacting fermions. In some cases it allows to obtain exact
solutions of the problems that are highly non-trivial in the fermionic
language. The canonical example of this is the Tomonaga-Luttinger model
equivalent to free bosons. Even in the case when bosonization does not
produce free bosons, it often constitutes a convenient starting point
for the development of the theory. Using this approach, effects of
backscattering \cite{ChuiLee,Voit}, impurities \cite{KaneFisher}, and
underlying periodic potential \cite{Voit1} on the LL were
explored. Many particular realizations of the LL states
were addressed, both theoretically and experimentally 
\cite{Giamarchi,Chang}. 

Nowadays, there is a growing interest in non-equilibrium phenomena in the
LL phase. In particular, in a recent experiment \cite{ChenDirks}, the tunneling
spectroscopy of a biased LL was carried out. A similar approach was
implemented to study experimentally carbon nanotubes \cite{Dirks} and
quantum Hall edges \cite{LeSueur,LeSueurAltimiras,Altimiras}. 
The experimental advances motivated theoretical interest to
quantum wires out of
equilibrium\cite{Jacobs07,Gutman1,Gutman3,Trushin08,Pugnetti09,NgoDinh10,Bena10,Takei10,Gutman4,Gutman11}. 
In a related line of research, non-equilibrium chiral 1D systems have been
studied theoretically
\cite{Chalker07,Levkivskiy08,Kovrizhin10,Schneider11},
mainly in application to experiments on quantum Hall edge-states interferometry.

In a series of papers by two of the authors and Gefen
\cite{Gutman1,Gutman3} the theory of non-equilibrium
LL was developed. The formalism developed in these works combined the
bosonization with the non-equilibrium Keldysh action approach.  It was
shown that the calculation of single-particle Green function reduces
to the evaluation of certain Fredholm determinants, analogous to those
encountered in the context of counting statistics
\cite{Levitov1} and non-equilibrium orthogonality catastrophe
\cite{Abanin04,Snyman07}.  
This allowed to obtain comprehensive results for observables related
to the fermionic single-particle Green function, including the
tunneling density of states, electron distribution function, and
Aharonov-Bohm signal.  More recently, the counting statistics of the
charge transfer in a non-equilibrium LL was studied in \cite{Gutman4}.  

However, a number of fundamental questions have remained open. In
particular, it is important to understand 
what kind of correlations (if any) are induced
by the interaction between the electrons coming out of an interacting
wire. In particular, are the left- and right-movers that have passed through
the wire correlated? We will see below that at equilibrium no such
correlations exist. On the contrary, in a non-equilibrium LL electrons
experience a specific type of relaxation. We will show that, in course
of this  relaxation process, non-trivial
correlations in the electronic distributions are built. To characterize
these correlations is the main task of this work.   

Our approach to the problem is complementary to the one of 
Ref.~\onlinecite{Gutman3}.  Using the operator formalism, 
we determine explicitly the many-body density matrix $\hat{\rho}$ of
the LL conductor in a stationary non-equilibrium state. 
When written in the bosonic coherent-state basis, this non-equilibrium
density matrix $\hat{\rho}$ has a form of a Fredholm determinant.
The density matrix carries full information about many-body
correlations in the non-equilibrium LL state. We find that 
the calculation of fermionic correlation functions  with density
matrix $\hat{\rho}$  is greatly simplified by
using a ``refermionization'' procedure. In this way, we can evaluate all
fermionic correlation functions, with results expressed in terms of 
Fredholm determinants.  Having developed our general formalism, 
we  employ it to study four-point
correlation functions of fermionic fields out of equilibrium and
explore the correlations in the occupation numbers of the outgoing
electrons induced by the interaction.  

The structure of the paper is as follows.
Section \ref{sec:model} contains a description of a 
model of a LL conductor out of equilibrium. 
In Sec.~\ref{sec:bosonization} we give a short review of
bosonization and refermionization  and introduce 
concepts needed in the main part of the paper.
In Sec.~\ref{sec:general} we obtain the many-body density matrix of a
LL. Further, we show how the bosonization-refermionization allows one to
obtain correlation functions of a non-equilibrium LL. We verify that
the results for the single-particle Green function agrees with that obtained
previously in Ref.~\onlinecite{Gutman3}. 
In Sec.~\ref{sec:out} we explore the four-point correlation functions
of fermions that emerged out of the LL wire. 
Our result reveals non-trivial correlations in fermionic occupation
numbers.  Section \ref{sec:conclusion} contains a summary of our
findings.  

\section{The model}
\label{sec:model} 
\begin{figure}
\includegraphics[width=220pt]{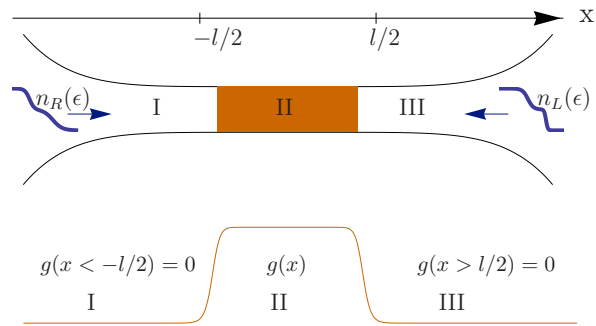}
\caption{\small (Color online). Schematic view a LL conductor driven
  out of equilibrium by the injection of non-equilibrium electrons
  (with distribution functions $n_{\eta}^{in}(\epsilon)$) from
  non-interacting leads.  
The leads are modeled via the assumption that the LL interaction
constant $g(x)$ is space dependent  
and vanishes outside the central part of the wire ($|x|<l/2$, region II).} 
\label{model}
\end{figure}

Let us specify our model of the non-equilibrium LL conductor. 
We consider a 1D wire populated by spinless electrons of two
chiralities (labeled by $\eta=R,L$) interacting via local
density-density interaction. 
The wire is  connected to non-interacting electrodes. To model the
later we will assume that the electron-electron interaction is  
switched off outside the central part of the wire (Fig. \ref{model})
so that the  Hamiltonian of the problem reads 
\begin{equation}
 H_{ee}=H_0+\frac12\int dx g(x) \left(\rho_{R}^2(x)+\rho_{L}^2(x)\right)\,.
\end{equation}
Here $H_0$ represents the kinetic energy part of the Hamiltonian;
$\rho_{R(L)}(x)$ is the density of right- (left-) moving electrons at
point $x$. The function $g(x)$ approaches a constant value deep inside
the interval $|x|<l/2$ (region II of Fig. \ref{model}) and is zero for
$|x|>l/2$ (regions I and III). This way of representing the
non-interacting leads was used extensively in the literature for the
analysis of the transport properties of Luttinger liquids
\cite{Maslov, Ponomarenko, Trauzettel}. 

The non-equilibrium state of the wire is induced by the injection of
the electrons from the leads. Following
Refs.~\onlinecite{Gutman3}, we assume that 
the right- moving electrons coming into
the wire from the  left lead have a stationary
non-equilibrium distribution function $n_{R}(\epsilon)$, while the
left-moving electrons coming from the right lead are characterized by
the distribution function $n_{L}(\epsilon)$. The simplest
non-equilibrium state arises when the leads are held at different
temperatures $T_R$, $T_L$ and different chemical potentials $\mu_R$,
$\mu_L$ so that the distribution functions of the incoming electrons
are of the Fermi-Dirac form 
\begin{equation}
 n_{\eta}(\epsilon)=\frac{1}{1+e^{\left(\epsilon-\mu_\eta\right)/T_\eta}}\,.
\end{equation}
Such a state was named the ``partial equilibrium'' in
Ref. \cite{Gutman3}. More complicated distribution functions  
$n_{\eta}$, e.g. double-step distributions, can be generated if one
assumes that the leads are diffusive conductors. We refer the reader
to the Ref. \cite{Gutman3} for the comprehensive discussion of the
possible experimental realizations of the non-equilibrium Luttinger
liquids.   
  
Throughout the paper we assume the absence of the electron
backscattering in the wire which  can be caused e.g. by impurities.  
In order to avoid the electronic backscattering also from the
boundaries of the interaction region we suppose that the function
$g(x)$ is smooth on the length scale of the Fermi wavelength. Withing
this limitation, the formalism developed in this work can be applied
to the case of arbitrary dependence of the interaction constant on
$x$.  
However, the most interesting situation arises when  boundaries of the
interaction region are sharp  on the scale of the typical wavelength
$l_{T^*}=V_F/T^*$ of the {\it bosonic} excitations. Here $V_F$ is the
Fermi velocity and $T^*$ is some characteristic width of the
distribution functions $n_{\eta}$ (e.g. $T^*\sim max(T_R, T_L)$ in the
case of the partial equilibrium). Under such circumstance, there is
significant scattering of the {\it bosonic excitations} on the
boundaries of the interaction region which is the primary source of
the effects we are going to discuss in this paper. For this reason  we
concentrate bellow on this ``sharp boundary'' limit and model the
$x$-dependence of the interaction constant by  
\begin{equation}
 g(x)=g\Theta\left(l/2-|x|\right)\,.
\label{g_rec}
\end{equation}

In Ref. \cite{Gutman3} the single particle correlation functions of
the model described above were investigate in great detail. While
within the interacting part of the wire  those correlation functions
are highly non-trivial  
due to the combined effect of interaction and non-equilibrium, outside
the interaction region the single particle Green functions are much
simpler, reflecting the free dynamics of the electrons. Specifically, 
the single particle Green functions of the right electrons  to the
right from the interacting region (region III) has a form of the Green
functions of free fermions with some energy distribution
$n^{out}_{R}(\epsilon)$. 
In the same way, the left-movers in the region I can be viewed as
free particles with energy distribution  
$n_L^{out}(\epsilon)$. The correlation functions of incoming
electrons, i.e. right-movers in the region I and left-movers in the
region III, are not affected by the interaction. 
The distribution functions $n_{\eta}^{out}$ differ
from $n_{\eta}$ (except for the case of complete thermal equilibrium),
signaling the redistribution  of the electrons over energies upon the
passage through the interacting region $II$.

Now an important question arises. Is this redistribution (which,
crudely speaking, can be referred to as relaxation, although the
resulting electronic distribution functions are not of the Fermi-Dirac
type) is the only effect of interaction on the outgoing fermions?  
In particular, are the left- and right-movers coming out of the wire
independent?  
Note that energy relaxation in our device is of a rather special
character. Due to the strong constraints imposed by the integrability
of the LL model, no energy relaxation can occur in a uniform LL. The
relaxation actually takes place at   the boundaries of the interacting
wire and involves the scattering of the bosonic excitations from one
chiral brunch to another. Since bosons are crudely speaking
electron-hole pairs, we may anticipate  that the relaxation should be
accompanied by the buildup of correlations in the distribution
functions of left- and right-electrons. We show below that this is
indeed the case and the density matrix of outgoing fermions shows
strong correlations.
In particular, 
the irreducible correlation functions of the form
$\langle\langle n_{\eta}(x, \epsilon) n_{\eta'}(x', \epsilon')\rangle
\rangle$ between the occupation numbers of the outgoing fermions are
non-zero. 
These emerging correlations are in the focus of the present work. We
stress that the correlations in question are specifically
non-equilibrium effect absent under the equilibrium conditions.  

We are now ready to present the formalism that we use in this work to
explore correlations in a non-equilibrium LL wire. 
We  start with a short  review of the  standard
bosonization approach (in the form of ``constructive
bosonization''\cite{Haldane, Delft}) in application to equilibrium
interacting fermions. 
We make particular emphasis  on constructions that will be 
subsequently  employed in the analysis of the non-equilibrium state.

\section{Bosonization and refermionization }
\label{sec:bosonization}
Let us consider the 1D fermionic system with the Hamiltonian
\begin{equation}
 H=\int dx \psi^+(x)\frac{\hat{p}^2-k_F^2}{2m}\psi(x)+V_{int}\,.
\end{equation}
Here $k_F$ is the Fermi momentum and $V_{int}$ represents the
four-fermion interaction. Within the Luttinger liquid model  
one linearizes the spectrum of the fermions near the Fermi points and
adds unphysical (but irrelevant at energies smaller than the Fermi
energy)  states below the bottom of the Fermi sea. This way of
treating  the interacting fermions is justified for the
description of low-energy properties of the system. At high
energies effects related to the curvature of the spectrum (not
included in the LL model) may become sizeable, see, in particular
Ref.~\onlinecite{Glazman}.  

With the bottom of the Fermi sea pushed down to minus infinity,  the
Hilbert space of the problem is spanned by the fermionic creation and
annihilation operators $a^+_{\eta, k}$ , $a_{\eta, k}$ labeled by
momentum $-\infty<k< +\infty$ (counted from $k_F$) and the chirality
$\eta=R, L$ (in formulas, we will also use the notation $\eta=1$ for
right-movers and $\eta=-1$ for left-movers). It is convenient to make
the fermionic momenta discrete, assuming that the system is placed on a
ring of large circumference $L$. The thermodynamic  limit
$L\rightarrow\infty$ is taken at the end. The physical fermionic field
is decomposed into the sum of left- and right-moving parts according
to 
($\Lambda\rightarrow \infty$ stands for the high energy cutoff)
\begin{eqnarray}
 \psi(x)\sim\psi_{R}(x)e^{ik_F x} +\psi_L(x)e^{-ik_F x}\,,\\
\psi_{\eta}(x)=\frac{1}{\sqrt{L}}\sum_{k}a_{k, \eta}e^{ikx-|k|/\Lambda}\,.
\end{eqnarray}

The Hamiltonian of the LL model acquires the form
\begin{equation}
 H=V_F\sum_{k, \eta}\eta k \left(a_{\eta, k}^+a_{\eta, k}-n^0_{\eta}(k)\right)
+\frac12\int dx g(x) (\rho_L(x)+\rho_R(x))^2\,.
\label{Lut_Hamiltonian}
\end{equation}
Here $n^0_{\eta}(k)=\Theta\left(-\eta k\right)$ are occupation
numbers of the left- and right-movers in the free-fermion ground
state; the interaction term was taken in the form of local
density-density interaction with space-dependent interaction constant
as discussed in the previous section.

The central role in the Luttinger-liquid model is played by the
Fourier components of fermionic densties
\begin{equation}
 \rho_{\eta, q}=\sum_{k}a^+_{\eta,k+q}a_{\eta, k}
\end{equation}
which can be used to construct the set of bosonic creation and
annihilation operators according to ($\eta q>0$): 
\begin{eqnarray}
b^+_{\eta , q}=
\sqrt{\frac{2\pi}{L |q|}}\rho_{\eta, q}\,, \qquad
b_{\eta , q}=\sqrt{\frac{2\pi}{L |q|}}\rho_{\eta, -q}\,,
 \\
\left[b_{\eta , q}, b^+_{\eta' ,
    q'}\right]=\delta_{\eta\eta'}\delta_{qq'}\,. 
\end{eqnarray}

Let us denote by $\left|N_R, N_L\right\rangle$ the state of the system
which is a filled Fermi sea with $N_{R(L)}$ extra particles in the
right (left) brunch, i.e the state  characterized by the distribution  
functions $n_{\eta, k}^{N_\eta}=\Theta\left(-\eta k+2\pi
  N_{\eta}/L\right)$. All these states are annihilated by $b_{\eta,
  q}$ and are vacuum states from the point of view of the bosons.  
Any other state of the fermions can be generated by the action of the
bosonic raising operators onto $\left|N_R, N_L\right\rangle$. 
Thus, the operators $b_{\eta,
  q}\,, b^+_{\eta, q}$ together with the particle number operators
$N_\eta$ and the Klein factors $F_{\eta}\,, F^{+}_{\eta}$ changing the
total number of  fermions of corresponding chirality form the complete
operator set. In particular, the free-fermion Hamiltonian $H_0$ can be
reexpressed in terms of bosons as 
\begin{equation}
 H_0=V_F\sum_{\eta, q}\Theta(\eta q)|q|b^+_{\eta, q}b_{\eta,
   q}+\frac{\pi V_F}{L}\sum_{\eta}N_{\eta}(N_{\eta}+1)\,, 
\label{H0_bosons}
\end{equation}
while the fermionic field operators are given by the famous
bosonization identity 
\begin{equation}
 \psi^+_{\eta}(x)=\sqrt{\frac{\Lambda}{4\pi}}e^{-i\varphi_{\eta}(x)}F_\eta^+
\label{psi+_initial}\,.
\end{equation}
The phase $\varphi_{\eta}(x)$ is related to the corresponding density
$\rho_{\eta}(x)$ via 
\begin{equation}
 \rho_{\eta}(x)=\frac{\eta}{2\pi}\partial_{x}\varphi_{\eta}(x)\,.
\label{rho_varphi}
\end{equation}
Explicitly, in terms of bosonic creation and annihilation operators
the density and phase fields read as follows,
\begin{eqnarray}
\varphi_{\eta}(x)=i\sum_{q}\Theta(\eta q)
\sqrt{\frac{2\pi}{L|q|}}\left(e^{-iq x}b^+_{q, \eta}-e^{iq x}b_{q,
    \eta}\right) 
+\frac{2\pi \eta}{L}N_{\eta} x\,,
\label{varphi_initial}\\
\rho_{\eta}(x)=\sum_{\eta, q}\Theta\left(\eta q\right)\sqrt{\frac{|q|}{2\pi L}}
\left(e^{-iq x}b^+_{\eta, q}+e^{iqx}b_{\eta, q}\right)+\frac{1}{L}N_{\eta}\,.
\label{rho_initial}
\end{eqnarray}
In Eqs. (\ref{H0_bosons}, \ref{varphi_initial}, \ref{rho_initial}) the
last terms involving the total number of fermions are usually referred
to as the "zero mode" contributions to the energy, bosonic phase and
the density.  
We keep them in the equations above in order to maintain the
completeness of our short presentation of the bosonization
technique. On the other hand, this contributions do not play any
significant role in the non-equilibrium effects we are going to
discuss below. For this reason, we omit the zero-mode terms
altogether in the subsequent formulas.

The power of the bosonization approach rests upon the fact that the
interacting fermionic Hamiltonian (\ref{Lut_Hamiltonian})  can be
reexpressed as a {\it quadratic} function of bosonic operators. Thus,
$H$ can be diagonalized by linear transformation of bosons.  
Bellow we will need explicit expressions for the the  bosons diagonalizing $H$.
To derive them, it is convenient to start with the equations of motion
for the density operators  
$\rho_{\eta}(x, t)$,
\begin{equation}
 \frac{\partial}{\partial t}\left(
\begin{array}{c}
\rho_{R}(x, t)\\
\rho_{L}(x, t)
\end{array}
 \right)+\frac{\partial}{\partial x}\left[
\left(
	\begin{array}{cc}
		V_F +\frac{g(x)}{2\pi} & \frac{g(x)}{2\pi}\\
		-\frac{g(x)}{2\pi} & -V_F-\frac{g(x)}{2\pi}
	\end{array}
	\right)\left(
\begin{array}{c}
\rho_{R}(x, t)\\
\rho_{L}(x, t)
\end{array}
 \right)\right]=0\,.
\end{equation}
We decompose the solution of the equation above into the normal modes
according to  
\begin{equation}
\rho_{\eta}(x, t)=\sum_{q, \eta'}
\Theta(\eta' q)\sqrt{\frac{|q|}{2\pi L}}\left(
 u^*_{\eta', |q|}(\eta, x)\widetilde{b}^{+}_{\eta',q}e^{i |q|V_F t}+
 u_{\eta', |q|}(\eta, x)\widetilde{b}_{\eta',q}e^{-i |q|V_F t}
\right)\,.
\label{rho_initial_gen}
\end{equation}
Here $\widetilde{b}$ and $\widetilde{b}^+$ are the new bosonic
operators, and 
$u_{\eta', |q'|}(\eta, q)$ are the coefficient functions. 
It is convenient to organize the latter into two-component columns 
 $u_{\eta, q}(x)=\left(u_{\eta, q}(x, R), u_{\eta, q}(x, L)\right)^T$
 satisfying the equation  
\begin{equation}
	-iV_F q u_{\eta, q}(x)+\frac{\partial}{\partial x}\left[\tau_z
          h(x)u_{\eta, q}(x)\right]=0\,, 
\label{equation_u}
\end{equation}
where $\tau_z$ is the third Pauli matrix  and 
\begin{equation}
 h(x)=\left(
	\begin{array}{cc}
		V_F +\frac{g(x)}{2\pi} & \frac{g(x)}{2\pi}\\
		\frac{g(x)}{2\pi} & V_F+\frac{g(x)}{2\pi}
	\end{array}
	\right).
\label{h}
\end{equation}
Note that in order to construct the
densty fields we need only the solutions of Eq. (\ref{equation_u})
with $q>0$. Thus, in the discussion of the properties of $u_{\eta,
  q}(x)$ we will assume  
that $q>0$. On the other hand, for $q<0$ one can define $u_{\eta,
  q}(x)$ via the complex conjugate 
\begin{equation}
 u_{\eta, q<0}(x)\equiv u^{*}_{\eta, -q}(x).
\end{equation}
 This choice is consistent with Eq. (\ref{equation_u}).

\begin{figure}
\includegraphics[width=400pt]{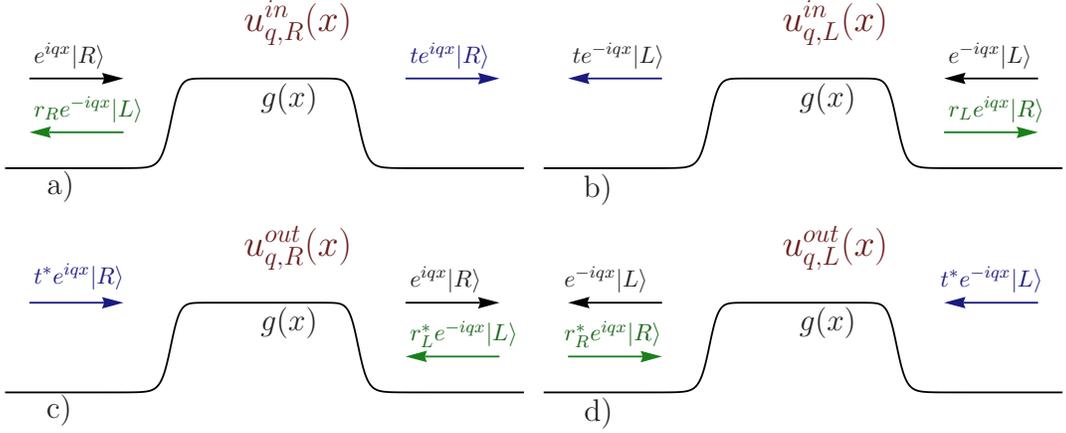}
\caption{\small (Color online). a), b) Bosonic  wave functions  of the in- scattering states outside the interacting part of the wire. c), d) Bosonic wave functions of the out- scattering states.   
}
\label{Scattering_states}
\end{figure}

Imposing different boundary conditions, one can construct two sets of
solutions of the Eq. (\ref{equation_u}). We denote them by ($q>0$):
$u_{q, \eta}^{in}$ and $u_{q, \eta}^{out}$.  
Far from the interacting region the solution $u^{in}_{q, R(L)}$
consits of a right (left) incoming wave and the waves scattered of the
interacting region. These are the $in$-states in the standard
terminology of the scattering theory. On the other hand $u^{out}_{q,
  R}$  and $u^{out}_{q, L}$ represent the $out$- scattering
states. They contain the outgoing wave and the waves incident on the
barrier. The asymptotic behavior of of $u_{\eta, q}^{in(out)}$ is
summarized on Fig. \ref{Scattering_states}. We have introduced the
bosonic $q$-dependent scattering coefficients of the interaction
region. Due to the time reversal symmetry the transmission amplitude 
$t_q$ for the right wave $u^{in}_{q, R}$ coincides with that of the
left wave, whereas reflection amplitudes for the right and left waves
are related via $r_R^*/r_L=-t^*/t$. If we assume the boundaries of the
interaction region to be sharp on the scale of the relevant
bosonic wavelength and adopt the model of the  rectangular-shaped
interaction constant (\ref{g_rec}) we find: 
\begin{eqnarray}
t_{q}&=&\frac{{\cal T}^2 e^{-i(1-K)ql}}{1-{\cal R}^2e^{2iK ql}}\,,\\
r_{R,q}&=&r_{L, q}=-2i{\cal R}\frac{e^{-i(1-K)q
                    l}}{1-{\cal R}^2e^{2iK q l}}\sin K q  l\,. 
\label{t_r}
\end{eqnarray} 
Here $K$ is the LL parameter of the interaction region, while ${\cal
  R}=(1-K)/(1+K)$ and ${\cal T}=\sqrt{1-{\cal R}^2}=2\sqrt{K}/(K+1)$ 
are reflection and  transmission amplitudes of  a single
interaction boundary.  

Clearly, outgoing solutions $u^{out}_{q, \eta}$  are not independent
of $u^{in}_{q, \eta}$. The two sets of solutions are related by the
elements of 
the bosonic scattering matrix of the interaction region  
$r_q$ and $t_q$:
\begin{eqnarray} 
u_{R, q}^{out}=t_{q}^* u_{R, q}^{in} + r^*_{L, q}u_{L, q}^{in}\,,\\
u_{L,q}^{out}=r_{R, q}^* u_{R, q}^{in}+t_{q}^* u_{L, q}^{in}\,.
\label{u_in_u_out}
\end{eqnarray}

Choosing $u_{\eta, q}^{in}(x)$ as the coefficient functions in
Eq. (\ref{rho_initial_gen}), we obtain the decomposition of the density
in terms of the in- bosons $b^{in}$ and $b^{+in}$, 
\begin{equation} 
\rho_{\eta}(x)=\sum_{q', \eta'}
\Theta(\eta' q')\sqrt{\frac{|q'|}{2\pi L}}\left(
 u^{in*}_{\eta', |q'|}(\eta, x)b^{+in}_{\eta',q'}+
 u^{in}_{\eta', |q'|}(\eta, x)b^{in}_{\eta',q'}
\right)\,.
\label{rho_initial_in}
\end{equation}
Comparing (\ref{rho_initial_in}) to (\ref{rho_initial}), one can read
off the relation of the initial and in-bosons: 
\begin{equation}
 b^{+in}_{\eta, q}=\sum_{\eta', q'}
\Theta(\eta' q')\left[u^{in}_{|q|,\eta}(q', \eta')b^{+}_{\eta', q'}-
u^{in}_{ |q|, \eta}(-q', \eta')b_{\eta', q'}
\right]\,.
\label{b_in}
\end{equation}
Operators $b^{in}_{\eta, q}$ are defined via hermitian conjugation and
$u^{in}_{q, \eta}(q')$ are the Fourier components of  
$u^{in}_{q, \eta}(x)$: 
\begin{equation}
 u^{in}_{q, \eta}(q', \eta')=\frac{1}{L}\sqrt{\frac{|q'|}{|q|}}\int dx
 e^{-iq'x}u^{in}_{q, \eta}(x, \eta')\,. 
\end{equation}
By construction, operators $b_{q, \eta}^{in}$ and $b_{q, \eta}^{+in}$
diagonalize the Hamiltonian $H$: 
\begin{equation}
 H=V_F\sum_{\eta, q}\Theta(\eta q) |q|b^{in+}_{\eta, q}b^{in}_{\eta,
   q}\,.
\label{H_b_in}
\end{equation}

The connection (\ref{rho_varphi}) of the phase fields $\varphi_{\eta}$
to the densities can now be used to derive  
the relation between $\varphi_{\eta}$ and the incoming bosons,
\begin{equation}
\varphi_{\eta}(x)=\frac{i}{V_F}\sum_{q', \eta'}
\Theta(\eta' q')\sqrt{\frac{2\pi}{L|q'|}}\left(
v^{in*}_{\eta', |q'|}(\eta, x)b^{+in}_{\eta',q'}-
v^{in}_{\eta', |q'|}(\eta, x)b^{in}_{\eta',q'}
\right)\,.
\label{varphi_initial_in}
\end{equation}
Here we have introduced the two-component columns $v^{in}_{\eta,q}(x)$ given by
(the matrix $h(x)$ was defined in (\ref{h}))
\begin{equation}
 v^{in}_{\eta',q'}(x)=h(x) u^{in}_{\eta', q'}(x)\,.
\label{v}
\end{equation}
Since $\psi^+_{\eta}(x)\sim \exp\left[-i\varphi_{\eta}(x)\right]$,
equation (\ref{varphi_initial_in})  
connects the fermionic field to the in-bosons. 

For the purposes of this work, it will be convenient to perform 
{\it refermionization} of  the in- bosons $b^{in}_{\eta,
  q}$ by introducing new fermionic fields 
$\psi_{\eta}^{in}(x)$ and their Fourier components $a_{\eta, k}^{in}$ according to
\begin{eqnarray}
\psi^{+in}_{\eta}(x)=\frac{1}{\sqrt{L}}\sum_{k}a^{+in}_{\eta,
  k}e^{-ikx}=\frac{1}{\sqrt{L\epsilon}}e^{-i\varphi^{in}(x)}F_\eta^+\,, 
\label{psi+_in}\\
\varphi^{in}_{\eta}(x)=i\sum_{q}\Theta(\eta q)
\sqrt{\frac{2\pi}{L|q|}}\left(e^{-iq x}b^{+in}_{q, \eta}- 
e^{iq x}b^{in}_{q, \eta}\right)\,.
\label{varphi_in}
\end{eqnarray}
The inverse relation reads
\begin{equation}
 b^{+in}_{\eta , q}=
\sqrt{\frac{2\pi}{L |q|}}\sum_{k}a^{+in}_{\eta,k+q}a^{in}_{\eta, k}\,, \qquad
b^{in}_{\eta , q}=\sqrt{\frac{2\pi}{L
    |q|}}\sum_{k}a^{+in}_{\eta,k-q}a^{in}_{\eta, k}\,. 
\label{b_in_via_a_in}
\end{equation}
Since the in- bosons solve the LL Hamiltonian, the in- fermions
$a_{\eta, k}^{in}$ are just free fermions with the time evolution 
\begin{equation}
 a^{in}_{\eta, k}(t)=a^{in}_{\eta, k}e^{-i\eta kv_F t}\,.
\label{a_in_eq_motion}
\end{equation}
Comparing Eqs. (\ref{varphi_initial_in}) and (\ref{varphi_in})
and taking into account the  
 boundary conditions imposed on $u^{in}_{\eta, q}$, we see that for
 $\eta x\rightarrow-\infty$ (e.g., for right fermions at
 $x\rightarrow-\infty$) the physical fermionic field $\psi_{\eta}(x)$
 is identical with $\psi^{i}_{\eta}(x)$\,,
\begin{equation}
 \psi_{\eta}(x)=\psi^{in}_{\eta}(x)\,,\qquad \eta x\rightarrow-\infty\,.
\label{psi_psi_in}
\end{equation}

So far, we were focussing on the in- solutions $u_{\eta, q}^{in}$. 
The out- solutions $u^{out}_{\eta, q}$ allow us to define the out-
bosons $b_{\eta, q}^{out}$ and fermions $a_{\eta, k}^{out}$ by 
equations  
analogous to (\ref{b_in}), (\ref{psi+_in}), and  (\ref{varphi_in}) up to
a replacement $in\rightarrow out$ everywhere. These out- fermions
and bosons also solve the LL Hamiltonian.  Just as the solutions
$u^{in}_{\eta, q}$ and $u^{out}_{\eta, q}$, the bosons $b_{\eta,
  q}^{in}$ and $b_{\eta, q}^{out}$ are related by the bosonic
transmission and reflection coefficients, 
\begin{eqnarray} 
b_{R, q}^{+out}=t_{q}^* b_{R, q}^{+in} + r^*_{L, q}b_{L, -q}^{+in}\,,\\
b_{L,q}^{+out}=r_{R, q}^* b_{R, -q}^{+in}+t_{q}^* b_{L, q}^{+in}\,.
\label{bosons_in_out}
\end{eqnarray}
 The out- fermions represent the physical fermions after the crossing
 of the interaction region in the sense that 
(cf. Eq.(\ref{psi_psi_in})): 
\begin{equation}
 \psi_{\eta}(x)=\psi^{out}_{\eta}(x)\,,\qquad \eta x\rightarrow +\infty\,.
\label{psi_psi_out}
\end{equation}

The refermionization  procedure described above turns out to be the
crucial step in the solution of the non-equilibrium LL model. The in-
fermions introduced in (\ref{psi+_in}) will provide us with the  
starting point for the construction of the non-equilibrium density
matrix of a LL wire. The out- fermions will be useful in our
discussion of the correlations among the electrons leaving the wire.

\section{Non-equilibrium Luttinger liquid}
\label{sec:general}

\subsection{Density matrix} 

In the previous section we have summarized the bosonization representation
of the LL Hamiltonian. 
Let us now turn to the analysis of the non-equilibrium LL in the setup
shown on Fig. \ref{model}. 
First, we need to understand what is the (many-body) density matrix
$\hat{\rho}$ describing the situation of Fig. \ref{model}. 
This density matrix should satisfy two conditions:\\
a) it should be stationary with respect to the LL Hamiltonian
(\ref{Lut_Hamiltonian});\\ 
b) any correlation function of the fermionic fields 
$\langle\psi_{\eta_1}(x_1, t_1)\ldots\psi^+_{\eta_i}(x_i,
t_i)\ldots\rangle$ evaluated with the density matrix  
$\hat{\rho}$ should be identical to the same correlation function of
free fermions with the density matrix 
\begin{equation}
 \hat{\rho}_0=\frac{1}{Z}\exp\left[-\sum_{\eta, k}\epsilon_{\eta}(k)
	\left(a_{\eta, k}^+a_{\eta, k}-n^0_{\eta}(k)\right)\right]
\label{density_matrix_initial}
\end{equation}
 as soon as all  coordinates of the fermionic fields  satisfy
 $\eta_ix_i<-l/2$. In Eq. (\ref{density_matrix_initial}) $Z$  stands
 for the normalization factor and the parameters  
$\epsilon_{\eta}(k)$ are determined by the distribution functions of
the fermions coming into the wire via 
\begin{equation}
 n_{\eta}(k)=\frac{1}{1+e^{\epsilon_{\eta}( k)}}\,.
\end{equation}

Both conditions a) and b) are fullfiled by the density matrix
\begin{equation}
 \hat{\rho}=\frac{1}{Z}\exp\left[-\sum_{\eta, k}\epsilon_{\eta}(k)
	\left(a^{+in}_{\eta, k}a^{in}_{\eta, k}-n^0_{\eta}(k)\right)\right]\,.
\label{density_matrix_in_fermions}
\end{equation}
The condition a) is satisfied due to the 
equation of motion (\ref{a_in_eq_motion}) for the incoming
fermions. The condition b) is satisfied due to the fact that the
fermionic fields $\psi_{\eta_i}(x_i)$ are just identical to
$\psi^{in}_{\eta_i}(x_i)$ as soon as  
$\eta_i x_i \rightarrow -\infty$.  

In equilibrium, $\epsilon_{\eta}(k)=\eta k V_F/T$, 
the density matrix in terms of bosons is simply given by 
\begin{equation}
 \hat{\rho}_{eq}\sim \exp\left[-\frac{V_F}{T}\sum_{q, \eta}\Theta(\eta
   q)|q|b^{+in}_{\eta, q}b^{in}_{\eta, q}\right]\,. 
\label{density_matrix_equilibrium}
\end{equation}
Together with Eq. (\ref{varphi_initial_in}) this implies that the
evaluation of the fermionic correlation functions in equilibrium is
reduced to the evaluation of the averages of the type ($g_{\eta}(q)$
and $g_{\eta}^*(q)$ are c-valued functions) 
\begin{equation}
Z[g^*_{\eta}, g_{\eta}]\equiv \left\langle \exp\left[\sum_{\eta,
      q}\Theta(\eta q)g_{\eta}(q)b^{in+}_{\eta, q} \right]  
	\exp\left[-\sum_{\eta, q}\Theta(\eta
          q)g^*_{\eta}(q)b^{in}_{\eta, q} \right]\right\rangle 
\label{prototypical_function}
\end{equation}
with the gaussian weight (\ref{density_matrix_equilibrium}), which is
a straightforward task. 

Translating the general non-equilibrium density matrix
(\ref{density_matrix_in_fermions}) into the bosonic
language is a much more non-trivial task.
As shown in Appendix \ref{Density_matrix_appendix}, 
it turns out to be possible to 
evaluate the matrix elements of $\hat{\rho}$ in the
basis of  bosonic coherent state  $|N, \beta\rangle$  (eigenstates of
the bosonic annihilation operators $b_{\eta, q}^{in}$) in the form of
one-dimensional Fredholm determinants.
Here we state only the
final result of this analysis, referring the reader to Appendix
\ref{Density_matrix_appendix} for the precise definitions and details
of the calculation:  
\begin{eqnarray}
	\langle \beta_R^*,\beta_L^* N_R, N_L|\hat{\rho}|N_R, N_L,
        \beta_R, \beta_L\rangle=D_R(\beta_R^*,\beta_R)
        D_L(\beta_L^*,\beta_L)\,, 
\label{R_final_ans_final_main}
\\
	D_{\eta}=\det\left[1- 
	e^{\Phi_{\eta}(x)}n_{\eta}^{N_{\eta}}(k) 
	e^{-\Phi_{\eta}(x)}(1-n_\eta(k))-
	e^{-\Phi^+_{\eta}(x)}(1-n_\eta^{N_\eta}(k))e^{\Phi^+_{\eta}(x)}n_{\eta}(k)
	\right]\,.
\label{determinants}
\end{eqnarray} 
In the determinants (\ref{determinants}) the distribution functions
$n_{\eta}(k)$ and $n^{N_{\eta}}_\eta(k)$ are considered as operators
diagonal in the momentum space. On the contrary, operators
$\Phi_{\eta}(x)$ are  diagonal with respect to the conjugate variable
$x$ and given by 
\begin{equation}
 \Phi_{\eta}(k_1-k_2)= \Theta(\eta(k_1-k_2))\sqrt{\frac{2\pi}{L
     |k_1-k_2|}}\beta_{\eta, k_1-k_2}\,.  
\end{equation}

The density matrix (\ref{R_final_ans_final_main}),
(\ref{determinants}), along with the
Hamiltonian (\ref{H_b_in}) and the expression
for the fermionic operators (\ref{varphi_initial_in}) contains the
whole information about the problem in the language of non-interacting
bosons. The natural next step is to calculate $n$-point fermionic
correlation functions (that determine various physical observables). 
This will be done in Sec.~\ref{s:corr_func}

\subsection{Electronic correlation function}
\label{s:corr_func}

Our goal now is to evaluate many-point fermionic Green functions that
have the form  (\ref{prototypical_function}), with the bosonic density matrix
given by
Eqs.~(\ref{R_final_ans_final_main}),(\ref{determinants}). While this can be
done by a direct calculation in the bosonic language, such a way turns
out to be quite tedious. A shorter way is to perform a
refermionization of the bosons $b^{in}$. Indeed,  according to
Eq.~(\ref{b_in_via_a_in}),  $b^{in+}$, $b^{in}$ are 
quadratic functions of the in-fermions  $a^{in+}$, $a^{in}$. 
Further, the density matrix is quadratic in terms of the in-fermions
as well, see Eq.~(\ref{density_matrix_in_fermions})
Thus,
the average   (\ref{prototypical_function}) can be expressed as trace  
of a certain operator, which is an {\it exponential of
a  quadratic form} with respect to fermions $a^{in}$. 
The evaluation of the trace
leads to (see Appendix \ref{trace} for details): 
\begin{eqnarray}
Z[g^*, g]=\Delta_{R}\left[\delta_R(x)\right]\Delta_L\left[\delta_L(x)\right]\,,
\label{prototypical_function_answer}
\\
\Delta_\eta[\delta_\eta(x)]=
\det\left[\left(1-n^0_{\eta}(k)+e^{-i\delta_\eta(x)}n^0_{\eta}(k)  
  \right)^{-1}\left( 1-n_\eta(k)+e^{-i\delta_\eta(x)}n_\eta(k)
  \right)\right]\,,
\label{Delta_Def}
\\
\delta_{\eta}(x)=i\sum_{q}\sqrt{\frac{2\pi}{L |q|}}\Theta(\eta q)\left(
g_{\eta, q} e^{iqx}-g^*_{\eta, q}e^{-iqx}
\right)\,.	
\label{delta_def}
\end{eqnarray}
In Eq.~(\ref{Delta_Def}) we can recognize a one-dimensional functional
determinant of the type discovered in \cite{Gutman3} in the
context of a single-particle Green function. 
The first factor in square brackets of Eq.~(\ref{Delta_Def})
involves the zero-temperature distribution function
$n^0_\eta(k)$ and serves as a regularization. It ensures that at zero
temperature $Z[g^*, g]$, which is an average of  a bosonic normal-ordered
expression, is identically equal to unity.   

Having derived the general result, we turn to evaluation of many-point
fermionic correlation functions.  Let
us consider the average of a product of $n$ operators
$\psi_{\eta_i}(x_i, t_i)\,, i=1\ldots n$ and $n$ operators
$\psi^+_{\eta_i}(x_i, t_i)\,,i=n+1\ldots 2 n$, 
\begin{equation}
 M_{\eta_1\ldots \eta_{2n}}(x_{1}, t_1, \ldots x_{2n},
 t_{2n})=\left\langle \psi_{\eta_{1}}(x_1, t_1)\ldots 
\psi_{\eta_n}(x_n, t_n)\psi^+_{\eta_{n+1}}(x_{n+1}, t_{n+1})\ldots
\psi^+_{\eta_{2n}}(x_{2n}, t_{2n})\right\rangle\,.  
\label{M_def}
\end{equation}
Representing the fermionic fields by bosonic exponets according to
(\ref{psi+_initial}, \ref{varphi_initial_in}) 
and brining the product of the exponents into the normal ordered form 
we get
\begin{eqnarray}
 M_{\eta_1\ldots \eta_{2n}}(x_{1}, t_1, \ldots x_{2n}, t_{2n})=
 M^0_{\eta_1\ldots \eta_{2n}}(x_{1}, t_1, \ldots x_{2n}, t_{2n})
Z[g^*, g]\,,
\label{M_form}
\\
	g_{\eta}(q)=-\frac1{V_F}\sqrt{\frac{2\pi}{L|q|}}
	\sum_{i=1}^{n}\zeta_i v^{in*}_{\eta,  |q|}\left( \eta_i, x_i
        \right)e^{i\eta q V_F t_i} 
	\,, \qquad
	g^*_{\eta}(q)=-\frac1{V_F}\sqrt{\frac{2\pi}{L|q|}}\sum_{i=1}^{n}\zeta_i
        v^{in}_{\eta, |q|} 
	\left( \eta_i, x_i \right)e^{-i\eta qV_F t_i}\,.
\label{g}
\end{eqnarray}
Here $\zeta_i=1$ for $i=1,\ldots n$, while $\zeta_i=-1$ for $i=n+1,\ldots 2n$. 
In the expression (\ref{M_form}) the first factor $M^0$ arises due to normal
ordering of the bosons and is nothing but  
the corresponding correlation function of $\psi$-operators in
zero-temperatures LL. This factor depends in particular on the
odering of $\psi$-operators in  (\ref{M_def}). On the other hand,  
the second factor in (\ref{M_form}) accumulates the effect of the
distribution functions of the incoming electrons
$n_{\eta}(\epsilon)$. This factor equals unity at zero temperature and
is the same for all the fermionic correlators which differ only by the
ordering of the Fermi fields.  
Explicit expression for the factor $Z[g^*_{\eta}, g_{\eta}]$ is given
by the Fredholm determinants (\ref{Delta_Def}) with the phases
$\delta_{\eta}(x)$ expressed in terms of the functions 
$v_{\eta, q}(x, \eta')$ via [cf. Eqs.~(\ref{delta_def},\ref{g})]
\begin{equation}
	\nabla_x \delta_{\eta}(x)=\frac{2\pi
          \eta}{V_F}\int\frac{dq}{2\pi}\sum_{i=1}^n\zeta_i 
	v^{in}_{\eta, q}\left( \eta_i, x_i \right)e^{-iqV_F t_i-i\eta q x}\,. 
\label{nabla_delta_ans}
\end{equation}
Equations (\ref{M_form}), (\ref{prototypical_function_answer}) and
(\ref{nabla_delta_ans}) together with (\ref{equation_u}, \ref{v})
provide the full solution of the non-equilibrium LL problem. It is
straightforward to check that the two-point (single-particle) 
correlation functions
found previously in Ref.\cite{Gutman3} are correctly reproduced. 

Expression for the phases $\delta_\eta(x)$ acquires particularly
simple form in the case when all the fermionic fields in (\ref{M_def})
are taken outside the interacting part of the wire, i.e. $|x_i|>l/2$
for all $i=1,\ldots 2n$. Taking into account the boundary conditions
imposed on the functions $u^{in}_{\eta, q}$ we get 
\begin{equation}
	\nabla_x \delta_{\eta}(x)=2\pi \eta \int\frac{dq}{2\pi}\sum_i\zeta_i
	e^{iq\left( \eta_i u_i -\eta x \right)}\left[ \Theta\left(
            \eta_i x_i \right)\left(  
	t_{q}\delta_{\eta,\eta_i}+r_{\eta,q}
	\delta_{\eta, -\eta_i}\right)+\Theta(-\eta_i
      x_i)\delta_{\eta\eta_i} \right]\,.
\label{nabla_delta_outside}
\end{equation}
Here we have introduced the light-cone coordinates $u_i=x_i-\eta_i V_F t_i$.
According to  (\ref{nabla_delta_outside}), in the case of the
electronic correlations  at the input of the central part of the wire
(i.e. for all $\eta_i x_i<-l/2$) the phase $\delta_{\eta}(x)$ are
completely independent from the properties of the interaction
region. This ensures the coincidence of the correlation functions with
that of the free fermions with the density matrix
(\ref{density_matrix_initial}). 
On the other hand, for the correlations  of the fermions at the output
of the interacting region (i.e. $\eta_i x_i>l/2$) the phases
$\delta_{\eta}(x)$ are determined by the bosonic transmition and
reflection amplitudes.

\section{Correlations in the outgoing fermions}
\label{sec:out}

\subsection{Single-particle Green functions}

We are now in a position to turn to 
the correlations between the electrons going out of the interacting
part of wire. The simplest quantities characterizing this correlations
are the four-point correlation functions of the type (\ref{M_def})
where the right fermions are ``measured'' to the right of the
interaction region, while the left electrons are ``measured'' to the
left, i.e. we consider the behavior of  Eq.~(\ref{M_def}) for $\eta_i
x_i>l/2$. Intuitively, the electrons which have left the interaction
region are just free electrons. We can make this statement
mathematically precise by recalling Eq. (\ref{psi_psi_out}) showing
that for $\eta_i x_i>l/2$ physical electronic fields coincide with the
fields of out- fermions, having just free {\it dynamics} 
\begin{equation}
 \psi_{\eta}(x,t)=\psi_{\eta}^{out}(x, t)=\psi_{\eta}^{out}(x-\eta V_F
 t)\,, \qquad \eta x>l/2\,.
\end{equation}
Thus, the question arises, if any of the correlation functions (\ref{M_def})
are non-trivial. However, apart from the dynamics, there is another
important ingredient of the correlation functions. This is the density
matrix.  

At equilibrium, the density matrix in terms of the in-bosons is
given by (\ref{density_matrix_equilibrium}). 
Using the interrelation (\ref{bosons_in_out}) of in- and out- bosons
and refermionizing the out-bosons, one  
readily finds
\begin{equation}
 \hat{\rho}_{eq}=\frac{1}{Z}\exp\left[-\frac1T\sum_{\eta, k}k V_F
	\left(a^{+out}_{\eta, k}a^{out}_{\eta, k}-n^0_{\eta}(k)\right)\right]\,.
\end{equation}
Thus, at equilibrium the out-electrons have both free dynamics and
trivial density matrix. 
In this situation none of the  correlation functions (\ref{M_def})
with $\eta_i x_i>l/2$ bears any trace of the interaction in the
central part of the wire. In particular, the prefactor $M^0$ entering
Eq.~(\ref{M_form})  for the correlation functions
at the output of the wire
is just the free fermion zero temperature correlation function. 

The situation  changes drastically in a non-equilibrium system. The
density matrix $\hat{\rho}$ becomes now a complicated function of the
out- fermions, incorporating non-trivial many-particle
correlations. It is important to distinguish between the
correlation functions which are determined solely by the dynamics
and the correlations also governed by the electronic distribution. The
retarded  
 and the Keldysh single particle Green functions are the simplest
 examples of the former and the latter.  
For the non-interacting electrons $G_{\eta}^{R0}(x, \tau)\sim
\delta(x-\eta V_F \tau)$. Thus,   
for the retarded function we immediately get
\begin{equation}
 G_{\eta}^{R}(x, \tau)=G_{\eta}^{R0}(x, \tau)\Delta_R[\delta_R\equiv
 0]\Delta_L[\delta_L\equiv 0]=G_{\eta}^{R0}(x, \tau)\,. 
\end{equation}
On the other hand, the Keldysh component of the Green function  (say
for right electrons)  
$G^K_R(x_R, \tau_R/2; x_R, -\tau_R/2)=-i[\psi_{R}(x_R,\tau_R/2),
\psi^+_{R}(x_R,-\tau_R/2)] 
\equiv G^K(\tau_R)$ receives non-trivial contribution from the
combined effect of interaction and non-equilibrium, 
\begin{equation}
 G_{R}^{K}(x_R, \tau_R/2, x_R, -\tau_R/2)=G_{R}^{K0}(0,
 \tau_R)\Delta_R[\delta_R]\Delta_L[\delta_L]\,,
\label{GK}
\end{equation}
where the phases $\delta_{\eta}$ are given by 
\begin{equation}
	\nabla_x \delta_{\eta}(x)=-4\pi i \eta \int\frac{dq}{2\pi}
	e^{iq (x_R-\eta x)}\sin\frac{q V_F \tau_R}{2}\left( 
	t_{q}\delta_{\eta,R}+r_{L,q}	\delta_{\eta, L}\right)\,.
\label{nabla_delta_outside_out}
\end{equation}
Our choice for the arguments of the fermionic fields in the definition
of the Keldysh Green function is somewhat over complicated, since
$G^K$ is actually independent of $x_R$ ( as long as
$x_R>l/2$). However we keep $x_R$ explicit for the convenience of the
forthcoming disscussion of the four-point correlations. 

\begin{figure}
\includegraphics[width=220pt]{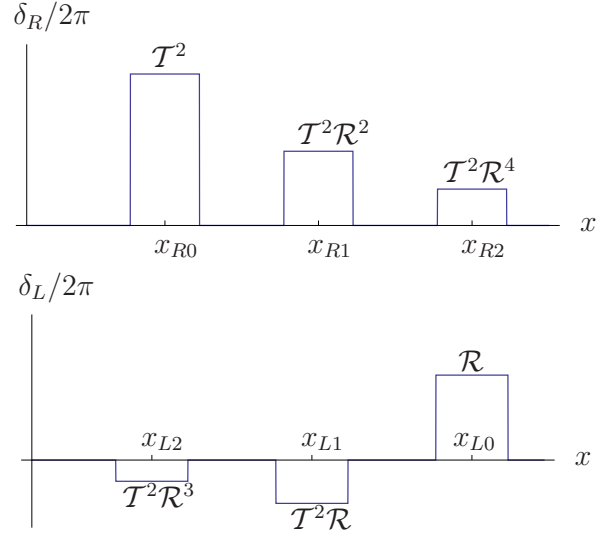}
\caption{\small The phases $\delta_{R}(x)$ and $\delta_L(x)$ governing
  the Keldysh Green function of the right-movers. The coordinates
  $x_{\eta,0}\,, x_{\eta,1}\ldots$ are given by
  Eqs. (\ref{x_Rm},\ref{x_Lm}).} 
\label{delta_pic}
\end{figure}

Exploiting the explicit expression (\ref{t_r}) for the  transmition
and reflection amplitudes in the  
``sharp boundary'' model we get the phases $\delta_{\eta}$ in the
characteristic form of sequence of rectangular pulses of length
$v_F\tau_R$ \cite{Gutman3}. The amplitudes of the pulses are shown on
Fig. \ref{delta_pic}. 
Their positions are given by  
\begin{equation}
 x_{Rm}=x_R-l+(2m+1)Kl \,, \qquad m=0\,,1\,,\ldots 
\label{x_Rm}
\end{equation}
for $\delta_{R}(x)$ and
\begin{equation}
 x_{Lm}=-x_R+l-2m Kl \,, \qquad m=0\,,1\,,\ldots 
\label{x_Lm}
\end{equation}
for the left phase. 
Analytically
\begin{equation}
 \delta_{\eta}(x)=2\pi \sum_{m=0}^{\infty}\alpha_{\eta, m}
 w_{\tau_R}(x-x_{\eta, m} )\,. 
\end{equation}
Here the coefficients
\begin{equation}
\alpha_{R, m}={\cal T}^2 {\cal R}^{2m}\,, \qquad \alpha_{L,m}= 
\left\{
\begin{array}{cc}
{\cal R}\,, & m=0\\
-{\cal T}^2 {\cal R}^{2m-1}\,, & m=1\,, 2\,,\ldots
\end{array}
\right.\,,
\label{alpha_def}
\end{equation}
and we have introduced
\begin{equation}
 w_{\tau}(x)= \Theta\left(\frac{|\tau|}{2}-|x|\right)\sign\tau\,.
\label{w_def}
\end{equation}

Equation (\ref{GK}) can be interpreted in terms of the  fermionic
occupation numbers at the output of the wire, 
\begin{equation}
1- 2 n^{out}_{\eta}(\epsilon)=iV_F\int d\tau_R e^{i\epsilon
  \tau_R}G^K_{\eta}(\tau_R)\,. 
\end{equation}
Since the $n^{out}_{\eta}(\epsilon)$ are different from the
distribution functions of the incoming electrons $n_{\eta}(\epsilon)$
(except for the equilibrium), electrons experience  relaxation upon
crossing the interaction region \cite{Gutman3}.

\subsection{Two-particle Green functions}
\label{s:two_particle_GF}

Let us now turn to a deeper characterization of the
density matrix for the outgoing electrons that is provided by the four-point
correlation functions. Particularly intriguing are interaction-induced
correlations between the left- and right-movers. To
reveal them we consider\cite{comment_arguments} 
\begin{equation}
 M_{RL}(x_R, \tau_R; x_L, \tau_L)=
\langle\left[ \psi_{R}\left( x_R, \frac{\tau_R}{2} \right), 
	\psi_{R}^+\left( x_R,-\frac{\tau_R}{2} \right) \right]
	\left[ \psi_{L}\left( x_L, \frac{\tau_L}{2} \right), 
	\psi_{L}^+\left( x_L, -\frac{\tau_L}{2} \right)
      \right]\rangle+G^K_R(\tau_R)G^K_L(\tau_L)\,. 
\label{MRL_def}
\end{equation}
were  $x_R>l/2$ and $x_L<-l/2$; the square brackets $[\cdot,\cdot]$
stand for the commutator. Note 
that (\ref{MRL_def}) is the only non-trivial irreducible correlation
function of two right and two left fields. Replacement of one of the
commutators in (\ref{MRL_def}) by anticommutator would immediately
lead to the decoupling of left and right  fermions under the
average. It is easy to see that the Fourier transform of
(\ref{MRL_def}) has the meaning of the irreducible correlator
of two fermionic distribution functions, 
\begin{equation}
 \langle\langle n_R(x_R, \epsilon_R) n_L(x_L,
 \epsilon_L)\rangle\rangle=\frac{V_F^2}{4}\int d\tau_R d\tau_L 
	 M(x_R, \tau_R; x_L, \tau_L)e^{i\epsilon_R \tau_R+i\epsilon_L\tau_L}\,.
\label{nRnL_def}
\end{equation}

Each of the  phases $\delta_{\eta}(x)$ corresponding to
(\ref{MRL_def}) is a sum as of a pulse sequence encountered previously
in the discussion of the Keldysh function $G^K_R$ and an analogous
contribution originating from the pair of left operators in
(\ref{MRL_def}). This last contribution consists of pulses of the
length $V_F\tau_L$ located at 
\begin{equation}
 x'_{Rm}=-x_L-l+2 m Kl \,, \qquad m=0\,,1\,,\ldots 
\end{equation}
for $\delta_{R}(x)$ and
\begin{equation}
 x'_{Lm}=x_L+l-(2m+1) Kl \,, \qquad m=0\,,1\,,\ldots 
\end{equation}
for the left phase. 

The dependence of (\ref{nRnL_def}) on the coordinates $x_{\eta}$ and
the ``optical length'' of the wire $K l$ can be easily understood. We
note that the scale $l_{T*}/V_F$ for the times $\tau_\eta$ supporting
non-zero correlation functions is set by the inverse characteristic
width of the energy distributions of the incoming electrons. Under the
assumption that the length of the interacting wire is large and $Kl$
exceeds $l_{T^*}$  we can neglect the  mutual effect of the
non-overlapping pulses in $\delta_{R(L)}$ and represent the
corresponding determinants $\Delta_{\eta}[\delta_{\eta}(x)]$ by a
product of determinants for individual pulses. One immediately
concludes that occupation numbers of left- and right- movers are
uncorrelated  unless the pulses coming from right and left operators
overlap. The overlap happens if  $X_{RL}=x_R+x_L$ is close to odd
multiple of $Kl$. We can now recast (\ref{nRnL_def}) into the form 
\begin{equation} 
 \langle\langle n_R(x_R, \epsilon_R) n_L(x_L, \epsilon_L)\rangle\rangle=
\sum_{ m \in \mbox{odd } }f^{RL}_m(\epsilon_R, \epsilon_L, X_{RL}-m Kl)\,.
\label{nRnL_form}
\end{equation}
The functions $f^{RL}_m(\epsilon_R, \epsilon_L, y)$ are  independent
of $l$ and decay on the distance $l_{T^*}$. 
They are given by
\begin{eqnarray}
f^{RL}_m(\epsilon_R, \epsilon_L, y)=-\frac{V_F^2}{4}\int d\tau_L
d\tau_L G^K_R(\tau_R)G^K_L(\tau_L) 
\left(A_R A_L-1\right)i\epsilon_R\tau_R+i\epsilon_L\tau_L\,,\\
A_R=\prod_{n=\max\left(0, \tilde{m}\right)}^{\infty}
\frac{\Delta_R\left[\alpha_{R, n-\tilde{m}}w_{\tau_R}(x)+\alpha_{L,
      n}w_{\tau_L}(x+y)\right]} 
{\Delta_R\left[\alpha_{R, n-\tilde{m}}w_{\tau_R}(x)\right]
\Delta_R\left[\alpha_{L, n}w_{\tau_L}(x)\right]}\,,
\label{AR}
\\
A_L=
\prod_{n=\max\left(0, \tilde{m}\right)}^{\infty}
\frac{\Delta_L\left[\alpha_{R, n-\tilde{m}}w_{\tau_L}(x-y)+\alpha_{L,
      n}w_{\tau_R}(x)\right]} 
{\Delta_L\left[\alpha_{R, n-\tilde{m}}w_{\tau_L}(x)\right]
\Delta_L\left[\alpha_{L, n}w_{\tau_R}(x)\right]}\,.
\label{AL}
\end{eqnarray}
Here $\tilde{m}=(m+1)/2$, and we use the notations introduced in
Eqs. (\ref{alpha_def}, \ref{w_def}).  

Similarly to (\ref{MRL_def}, \ref{nRnL_def}), one can define the
irreducible correlation function of the occupation numbers of the
right-movers: 
\begin{eqnarray} 
\langle\langle n_R(x_R, \epsilon_R) n_R(x_R',
\epsilon_R')\rangle\rangle=\frac{V_F^2}{4}\int d\tau_R d\tau_R'  
M_{RR}(x_R, \tau_R; x_R', \tau_R')e^{i\epsilon_R \tau_R+i\epsilon_R'\tau_R}\,, 
\label{nRnR_def}
\\
M_{RR}(x_R, \tau_R; x_R', \tau_R')=
\langle\left[ \psi_{R}\left( x_R, \frac{\tau_R}{2} \right), 
	\psi_{R}^+\left( x_R,-\frac{\tau_R}{2} \right) \right]
	\left[ \psi_{R}\left( x_R', \frac{\tau_R'}{2} \right), 
	\psi_{R}^+\left( x_R', -\frac{\tau_R}{2} \right)
      \right]\rangle+G^K_R(\tau_R)G^K_R(\tau_R')\,. 
\label{MRR_def}
\end{eqnarray}
In order to be able to interpret (\ref{nRnR_def}) as the correlator of
the  distribution functions, we assume below that the two pairs of
$\psi$-operators in (\ref{MRR_def}) are inserted far from each other,
so that $X_{RR}\equiv 
x_R-x_R'\gg l_{T^*}$.  Under this assumption, just as in the case
of $R$-$L$ correlations, (\ref{MRR_def}) is the only one non-trivial
correlation function of four right electronic fields. Consideration
of phases $\delta_{\eta}(x)$ corresponding to (\ref{MRR_def}) shows
that the (\ref{nRnR_def}) is non-zero  provided that  
$X_{RR}\approx 2 Kl m$ and can be decomposed as (we set $\tilde{m}=m/2$):
\begin{eqnarray} 
 \langle\langle n_R(x_R, \epsilon_R) n_R(x_R', \epsilon_R')\rangle\rangle=
\sum_{m\in \mbox{even}\,, m\neq 0}f^{RR}_m(\epsilon_R, \epsilon_L, X_{RR}-mKl)\,,
\label{nRnR_form}\\
f^{RR}_m(\epsilon_R, \epsilon_L, y)=-\frac{V_F^2}{4}\int d\tau_R
d\tau_R' G^K_R(\tau_R)G^K_L(\tau_L) 
\left(A_RA_L-1\right)e^{i\epsilon_R\tau_R+i\epsilon_R'\tau_R'}\,,
\\
A_R=\prod_{n=\max\left(0, \tilde{m}\right)}^{\infty}
\frac{\Delta_R\left[\alpha_{R, n-\tilde{m}}w_{\tau_R}(x)+\alpha_{R,
      n}w_{\tau_R'}(x+y)\right]} 
{\Delta_R\left[\alpha_{R, n-\tilde{m}}w_{\tau_R}(x)\right]
\Delta_R\left[\alpha_{R, n}w_{\tau_R'}(x)\right]}\,,
\label{ARp}
\\
A_L=
\prod_{n=\max\left(0, \tilde{m}\right)}^{\infty}
\frac{\Delta_L\left[\alpha_{L, n-\tilde{m}}w_{\tau_R}(x)+\alpha_{L,
      n}w_{\tau_R'}(x-y)\right]} 
{\Delta_L\left[\alpha_{L, n-\tilde{m}}w_{\tau_L}(x)\right]
\Delta_L\left[\alpha_{L, n}w_{\tau_R'}(x)\right]}\,.
\label{ALp}
\end{eqnarray} 
The term with $m=0$ is omitted from the summation due to our
assumption that $X_{RR}>l_{T^*}$. 
Again, investigation of the functions $f^{RR}_m(\epsilon_R,
\epsilon_L, x)$ requires evaluation of the functional determinants
$\Delta_{\eta}$.  

The physics of obtained correlations is discussed in Sec.~\ref{s:discussion}.

\subsection{Discussion}
\label{s:discussion}

We assert that the correlations discovered in
Sec.~\ref{s:two_particle_GF} represent a quantum interference effect. 
To clarify this point, we make a digression and 
consider a very simple case of one
right boson with momentum $q$ populating the wire. This situation is
described by the density matrix 
\begin{equation}
 \hat{\rho}_{1b}=b^{+in}_{R,q}|0\rangle\langle 0|b^{in}_{R, q}\,,
\label{rho_1b}
\end{equation}
where $|0\rangle$ is the ground state. Using the connection of the in-
and out-bosons and then performing the refermionization  we translate
$\rho_{1b}$ into the out-fermions 
\begin{equation}
 \hat{\rho}_{1b}=\frac{2\pi}{L q}\sum_{k, k'}
\left( t_{q} a^{+out}_{R, k+q}a^{out}_{R, k}+r_{R, q}a^{+out}_{L,
    k-q}a^{out}_{L, k}\right) 
|0\rangle\langle 0|\left( t^*_{q} a^{+out}_{R, k'-q}a^{out}_{R,
    k}+r^*_{R, q}a^{+out}_{L, k'+q}a^{out}_{L, k}\right)\,. 
\label{den_matrix_simple}
\end{equation}
We can  now evaluate the correlation function of the left and right
distributions directly in the fermionic language and get 
\begin{multline}
 \langle\langle n_R(x_R, \epsilon_R) n_L(x_L, \epsilon_L)\rangle\rangle=
-\delta n_{R}(\epsilon_R)\delta n_L(\epsilon_L)\\+\frac{2\pi}{L q}
\left(n^0_{R+}-n^0_{R-}\right)\left(n^0_{L+}-n^0_{L-}\right)
\left(r^*_{R,q} t_{q} e^{iq(x_R+x_L)}+r_{R,q} t^* e^{-iqX_{RL}}\right)\,.
\label{cor_simple_ans}
\end{multline}
Here $\omega=q V_F$ is the frequency of the boson in the system;
$n^0_{\eta\pm}=\Theta(-(\epsilon_\eta\pm\omega/2))$  and  
\begin{eqnarray}
\delta n_{R}(\epsilon_R)\equiv
n^{out}_R(\epsilon_R)-n^0(\epsilon_R)=\frac{2\pi}{Lq}|t_{q}|^2 
\left(2n^0(\epsilon_R)-n^0(\epsilon_R+\omega)-n^0(\epsilon_R-\omega)\right)\,,\\
\delta n_{L}(\epsilon_L)\equiv
n^{out}_L(\epsilon_L)-n^0(\epsilon_R)=\frac{2\pi}{Lq}|r_{R,q}|^2 
\left(2n^0(\epsilon_L)-n^0(\epsilon_L+\omega)-n^0(\epsilon_L-\omega)\right)\,.\\
 \end{eqnarray}
The two terms in (\ref{cor_simple_ans}) have distinct physical origin. 
The first one originates from the probabilistic nature of the boson
transmission-reflection process. It would remain unchanged upon the
replacement  
of the density matrix by the statistical mixture 
\begin{equation}
\tilde{\hat{\rho}}_{1b}=
|t_{q}|^2 b^{+out}_{R, q}|0\rangle\langle 0|b^{out}_{R, q}+
|r_{R, q}|^2 b^{+out}_{L, -q}|0\rangle\langle 0|b^{out}_{L,- q}\,.
\label{den_matrix_simple_class}
\end{equation}
This term favors {\it anti-correlations} of electrons at energies of
equal signs.  This is a typical anti-correlation between the mutually
exclusive events. Indeed, suppose the boson we have injected into the
system was transmitted through the interacting part of the wire. In
this case at the output of our device there is a particle-hole
excitation in the right branch, i.e. a right electron at positive
energy and a right hole at negative energy. At the same time the left
branch has no excitation. On the contrary, if the boson was reflected
there  is  a left electron at positive energy and a left hole. The
right branch is empty.  
Thus, in such a probabilistic description there is no way to have an
excited electron both in the left and right branches, which results in
anti-correlations between the fermionic occupation numbers at energies
of the same sign. 

Note that the probabilistic contribution to the correlator does not
depend on the coordinates where the  
the occupation numbers are measured. This is quite general. Suppose we
have some complicated density matrix $\hat{\rho}$ in terms of
in-bosons. We translate it into the out-bosons. It is easy to see that
if we now switch to the classical description of the system by
dropping all the non-diagonal elements of the density matrix
(cf. transition from  
(\ref{den_matrix_simple}) to (\ref{den_matrix_simple_class})) we will
end up with the correlator of the occupation numbers independent on
$X_{RL}$. Only the matrix elements of $\hat{\rho}$ involving momentum
transfer from left to right branch can provide such a dependence.    

The second contribution to the correlator (\ref{cor_simple_ans})   is
due to the fact that  on the quantum level the scattering of a boson
creates {\it coherent} superposition of the state with a boson in the
right branch and a state with a boson in the left branch. 
Thus, the exited particle-hole pair is (virtually) simultaneously
present in {\it both} branches. To elucidate the  
effect of this quantum term, let us consider a slightly more general
density matrix which is a statistical mixture  
of (\ref{rho_1b}):
\begin{equation}
  \hat{\rho}_{mix}=\sum_{q}c_qb^{+in}_{R,q}|0\rangle\langle
  0|b^{in}_{R, q}\,, \qquad \sum_{q}c_q=1\,. 
\end{equation}
Now the quantum contribution is modified accordingly,
\begin{equation}
 \langle\langle n_R(x_R, \epsilon_R) n_L(x_L,
 \epsilon_L)\rangle\rangle_{quantum}= 
-4{\cal T}^2 {\cal R}\sum_{q}c_q\frac{2\pi}{L q}
\left(n^0_{R+}-n^0_{R-}\right)\left(n^0_{L+}-n^0_{L-}\right)
\frac{\sin Kql\sin qX_{RL}}{\left|1-{\cal R}^2 e^{2iKql}\right|^2}\,.
\label{cor_mix_ans}
\end{equation}
We have used here the explicit expressions (\ref{t_r}) for the
transmission and reflection amplitudes in the sharp-boundary model.  
Let the coefficients $c_q$ be peaked at some $q=q_0$. Under the
assumption that the peak width is much larger than $1/Kl$ and the
energies $\epsilon_{R(L)}$ are not too close to $q_0 V_F/2$, we can
average the expression under the sum over fast oscillations
oscillations on the scale $1/Kl$. The result is non-vanishing only
if  
$X_{RL}$ is close to odd multiple of $Kl$, $X_{RL}=m Kl +x$, $m\in
odd$, in which case 
\begin{equation}
 \langle\langle n_R(x_R, \epsilon_R) n_L(x_L,
 \epsilon_L)\rangle\rangle_{quantum}= 
-\frac{4\pi {\cal T}^2 {\cal R}^m\sign m}{1+{\cal
    R}^2}\sum_{q}c_q\frac{2\pi}{L q} 
\left(n^0_{R+}-n^0_{R-}\right)\left(n^0_{L+}-n^0_{L-}\right)\cos q x\,.
\label{cor_mix_ans_final}
\end{equation}
This result is in agreement with the general coordinate dependence
(\ref{nRnL_form}) of the correlator of fermionic distributions.

The density matrices $\hat{\rho}_{1b}$ and $\hat{\rho}_{mix}$ are very
simple as they contain just one fermionic excitation.
The truly non-equilibrium density matrix
(\ref{density_matrix_in_fermions}), when written in terms of bosons, is
much more complicated. It contains infinitely many terms representing
multiple-boson processes. To collect properly their contributions into
the correlation function  of left and right occupation numbers is the
task accomplished by the functional determinants
$\Delta_{\eta}[\delta_{\eta}(x)]$.   In general, both
classical noise and quantum interference contribute to the result.
The lesson we learnt from the analysis of simple density matrices
allows us to identify clear manifestations of quantum effects. 
First, this is the dependence of the correlation functions
on the coordinates, see Eq.~(\ref{nRnL_form}). Second, these are
positive correlations between distribution functions of left and right
movers at energies of the same sign. We will see in
Sec.~\ref{s:partial_equilibrium} below that the correlation functions
of a non-equilibrium LL do show both these features.

\subsection{Partial equilibrium}
\label{s:partial_equilibrium}

A detailed analysis of the functions $f^{RL}_m$ describing the
correlations in electronic distributions requires a careful
investigation of the determinants $\Delta_{\eta}[\delta_{\eta}(x)]$
with given distributions of the incoming electrons. 
In general, these  determinants can not be evaluated
analytically, except for asymptotic long-time behavior\cite{Gutman11} 
(which is not
sufficient for our purposes here). Thus, one has to resort to numerics
in order to achieve the comprehensive understanding  of the
correlations in the fermionic occupation numbers. 

Relegating such a numerical analysis for a future work, we focus below
on the case of partial equilibrium that can be treated analytically.  
This is the situation when  
distribution functions of the incoming fermions are of the Fermi-Dirac
form but with different temperatures  
$T_R$ and $T_L$ of the left- and right-movers\cite{comment2}. 
The functional determinants $\Delta_{\eta}[\delta_{\eta}(x)]$ become
under such circumstance exponentially quadratic  
functionals of phases \cite{Gutman3},
\begin{equation} 
 \ln\Delta_{\eta}=-\frac{1}{4\pi}\int_0^\infty\frac{dq}{2\pi}q\left(
   B_{\eta}\left( q V_F \right)-1 \right)|\delta_\eta(q)|^2 
\label{delta_partial_eq}\,,
\end{equation}
where $B_\eta(\omega)=\coth \omega/2T_\eta$ are the bosonic
distribution functions.  

One can now use (\ref{delta_partial_eq}) to evaluate the products of
the determinants (\ref{AR},\ref{AL},\ref{ARp},\ref{ALp})
analytically. Alternatively, we can apply Eq. (\ref{delta_partial_eq})
to the full functional determinants governing the correlations in the
electronic distributions.  
The Fourier components of the phases $\delta_{\eta}$ are easily read
off from (\ref{nabla_delta_outside_out}). 
For example, in the case of of the correlator of left and right
distribution functions  we have 
\begin{eqnarray}
 |\delta_{R}(q)|^2=\frac{(4\pi)^2}{q^2}\left[ 
	|t_q|^2\sin^2\frac{qV_F\tau_R}{2}+|r_q|^2\sin^2\frac{qV_F\tau_L}{2}+
	2it_{q}r_{q}^*\sin\frac{qV_F\tau_R}{2}\sin\frac{qV_F\tau_L}{2}\sin
        q(x_R+x_L) 
	\right]\,,\\
	\hspace{-2cm}
	|\delta_{L}(q)|^2=
	\frac{(4\pi)^2}{q^2}\left[ 
	|t_q|^2\sin^2\frac{qV_F\tau_R}{2}+|r_q|^2\sin^2\frac{qV_F\tau_L}{2}-2i
	t_{q}r_{q}^*\sin\frac{qV_F\tau_R}{2}\sin\frac{qV_F\tau_L}{2}\sin
        q(x_R+x_L) 
	\right]\,.
	\qquad
\end{eqnarray}
Plugging this into (\ref{delta_partial_eq}), averaging the expression
under the integral over fast oscillations  
of the bosonic transmition and reflection amplitudes (which is
equivalent to the neglect of the interference of non-overlapping
pulses in $\delta_{\eta}(x)$ discussed in the previous section) and
using the standard equality 
\begin{equation}
 -\int_{0}^{+\infty}\frac{dx}{x}(1-\cos \alpha
 x)\left(\coth\frac{x}{2}-1\right)=\ln \frac{\pi \alpha}{\sh\pi\alpha}\,, 
\end{equation}
one finds that the correlation function of left and right occupation
numbers has indeed the form  
(\ref{nRnL_form}) with
\begin{equation}
f^{RL}_m(\epsilon_R, \epsilon_L, x)=
-\frac{V_F^2}{4}\int d\tau_R d\tau_L 
	e^{ i\epsilon_L\tau_L+i\epsilon_R\tau_R }G^K_{R}(\tau_R)G^K_{L}(\tau_L)
	\left[\left(\frac{1-\frac{\ch\pi T_R\left( \tau_R+\tau_L
                \right)}{\ch2\pi T_R x}} 
	{1-\frac{\ch\pi T_L\left( \tau_R+\tau_L \right)}{\ch2\pi T_L x}}
	\frac{1 -\frac{\ch\pi T_L\left( \tau_R-\tau_L \right)}{\ch2\pi T_L  x}}
	{1-\frac{\ch\pi T_R \left( \tau_R-\tau_L \right)}{\ch2\pi T_R
            x}}\right)^{\gamma_m}-1 
	\right]\,.
\label{nRnL_ans}
\end{equation}
Here the exponents $\gamma_m$ are given by 
$\gamma_m=\sign m {\cal T}^2 {\cal R}^{|m|}/(1+{\cal R}^2)$, and the  
Keldysh Green functions of the outgoing fermions are \cite{Gutman3}
\begin{eqnarray}
 G^K_R(\tau_R)=\frac{1}{\pi V_F\tau_R}
	\left[\frac{\pi T_R \tau_R}{\sh \pi T_R \tau_R
          }\right]^{\frac{{\cal T}^2}{1+{\cal R}^2}} 
	\left[\frac{\pi  T_L\tau_R}{\sh \pi T_L \tau_R}
        \right]^{\frac{2{\cal R}^2}{1+{\cal R}^2}}\,,\\ 
 G^K_L(\tau_L)=\frac{1}{\pi V_F \tau_L}
	\left[\frac{\pi T_L \tau_L}{\sh \pi T_L \tau_L
          }\right]^{\frac{{\cal T}^2}{1+{\cal R}^2}} 
	\left[\frac{\pi  T_R\tau_L}{\sh \pi  T_R\tau_L
          }\right]^{\frac{2{\cal R}^2}{1+{\cal R}^2}}\,.\\ 
\end{eqnarray}

\begin{figure}
\includegraphics[width=250pt]{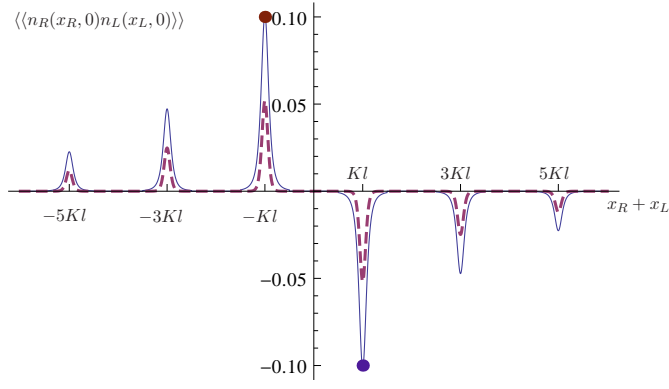}\caption{\small (Color
  online). The dependence of the correlator  
$\langle\langle n_{R}(x_R, 0)n_{L}(x_L, 0)\rangle\rangle$ on the sum
of coordinates $x_R+x_L$. Solid line corresponds to $T_R/T_L=10$ while
the dashed line is the result for $T_R/T_L=2$. Energy dependence of the correlator of left and right distribution functions at $x_R+x_L=Kl$ (blue dot) and $x_R+x_L=-Kl$ (brown dot) is shown on Fig. \ref{Max_cor_RL}.}
\label{RLX_dep}
\end{figure} 
\begin{figure}
\includegraphics[height=170pt]{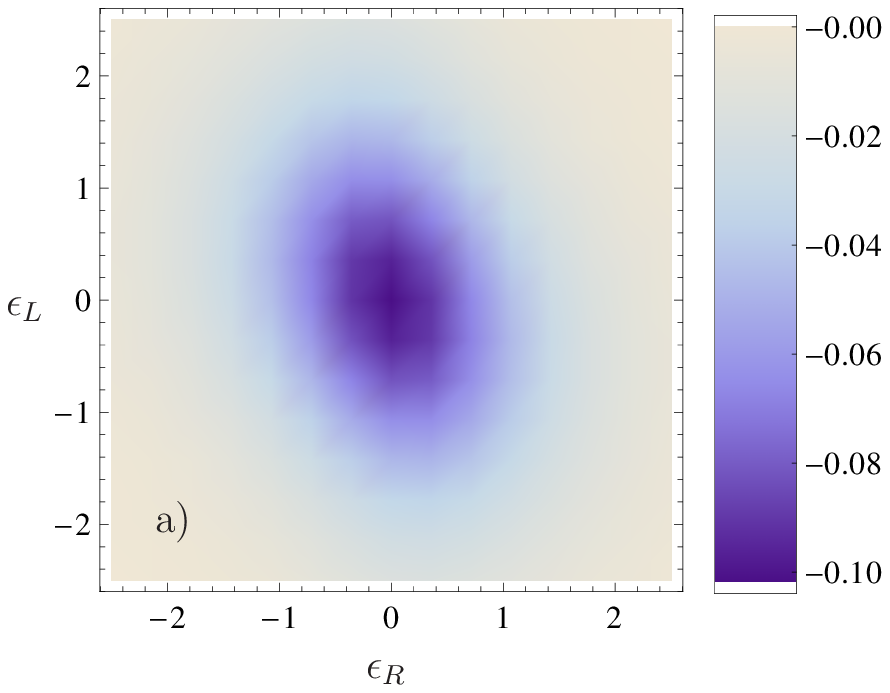}\hspace{1cm}
\includegraphics[height=170pt]{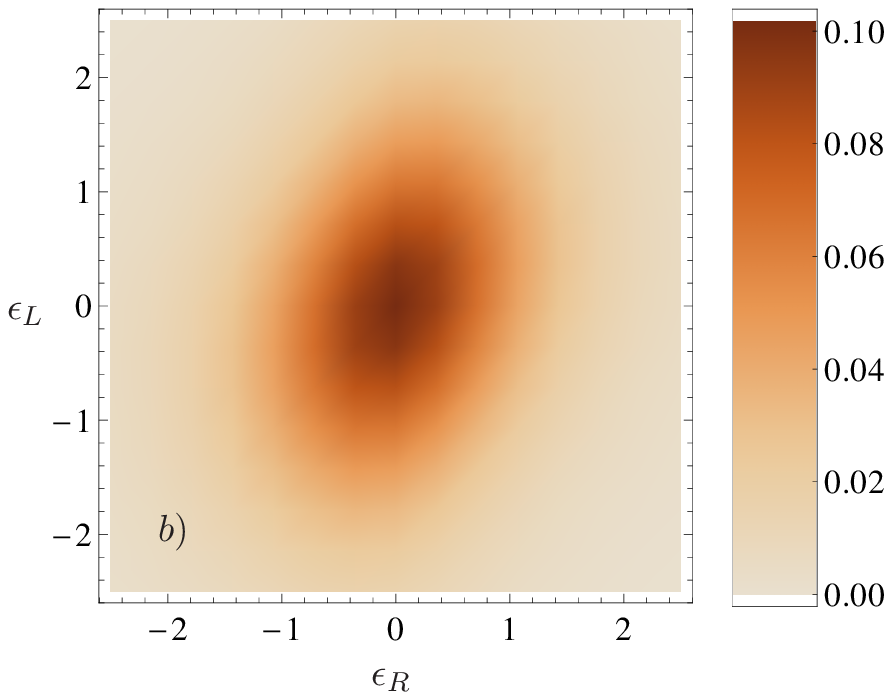}
\caption{\small (Color online).  Energy dependence of functions
  $f^{RL}_{1}(\epsilon_R, \epsilon_L, 0)$ and $f^{RL}_{-1}(\epsilon_R, \epsilon_L, 0)$ (plots (a) and (b)
  respectively). All energies are measured in units of the largest
  temperature $T_R=10 T_L$.     
}
\label{Max_cor_RL}
\end{figure}

\begin{figure}
\includegraphics[width=220pt]{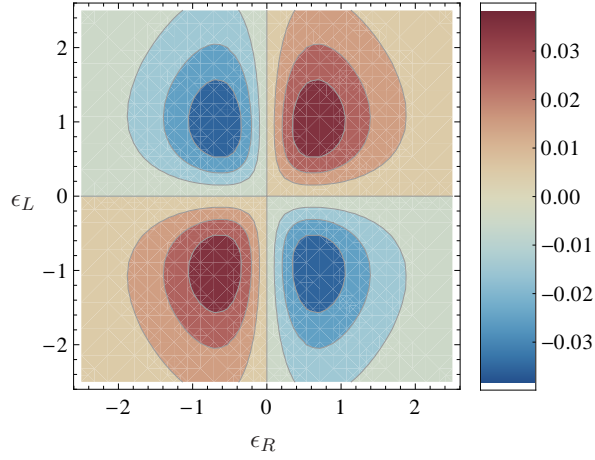}
\caption{\small (Color online). Correlator of the  distribution
  functions of left and right electrons integrated over the  
$x_R+x_L$ (see  Eq. (\ref{heps_def})).  The ratio of temperatures was taken
to be $T_R/T_L=10$.  
}
\label{Int_cor_RL}
\end{figure}

The Fourier transformation in (\ref{nRnL_ans})  can be evaluated
numerically. The results are exemplified on  
Fig. \ref{RLX_dep}, \ref{Max_cor_RL}  and \ref{Int_cor_RL}. To produce
the graphs we have chosen the LL parameter $K=0.2$ such that the
transmission and reflection amplitudes ${\cal T}$, ${\cal R}$ are
approximately equal. Figure \ref{RLX_dep} shows the zero energy
correlator  
$\langle\langle n_{R}(x_R, 0)n_{L}(x_L, 0)\rangle\rangle$ in its
dependence on $x_R+x_L$.  
Solid and dashed lines correspond to strong ($T_R/T_L=10$) and
comparatively weak ($T_R/T_L=2$) non-equilibrium.  
One observes the characteristic peaks  at
$x_R+x_L=(2m+1)Kl$. Occupation numbers of left and right movers are
anti-correlated for $x_R+x_L>0$ and correlated at $x_R+x_L<0$ (of
course, the situation will be reversed if one assumes that $T_R<T_L$).

Figure (\ref{Max_cor_RL}) demonstrates the  functions $f^{RL}_{\pm
  1}(\epsilon_R,\epsilon_L, 0)$. The correlations reach maximum near
$\epsilon_R=\epsilon_L=0$. In the case of $f^{RL}_{1}$ the
anti-correlations are somewhat more extended in the direction
$\epsilon_R=-\epsilon_L$ while the correlations in  
$f^{RL}_{-1}$ prefer to develop for $\epsilon_R$ and $\epsilon_L$ of
equal sign. This fact leads to a remarkable structure in the ``mean
correlations'' of the distribution functions of left- and right-
movers  as given by  
\begin{equation}
 h(\epsilon_R,\epsilon_L)\equiv \frac{T_R}{V_F}\int
 d(x_R+x_L)\langle\langle n_{R}(x_R, \epsilon_R)n_{L}(x_L,
 \epsilon_L)\rangle\rangle\,. 
\label{heps_def}
\end{equation}
(The prefactor was introduced to make $h$ dimensionless.)  The function
$h(\epsilon_R,\epsilon_L)$ is shown on Fig. \ref{Int_cor_RL}.  We
observe that on average the left- and right- movers are correlated
when their energies have equal signs and anti-correlated in the
opposite case. Note that this result is opposite to the one that
would be except on the basis of the  classical consideration of the bosonic
transmission-reflection process, see 
Sec.~\ref{s:discussion}. 

\begin{figure}
\includegraphics[height=170pt]{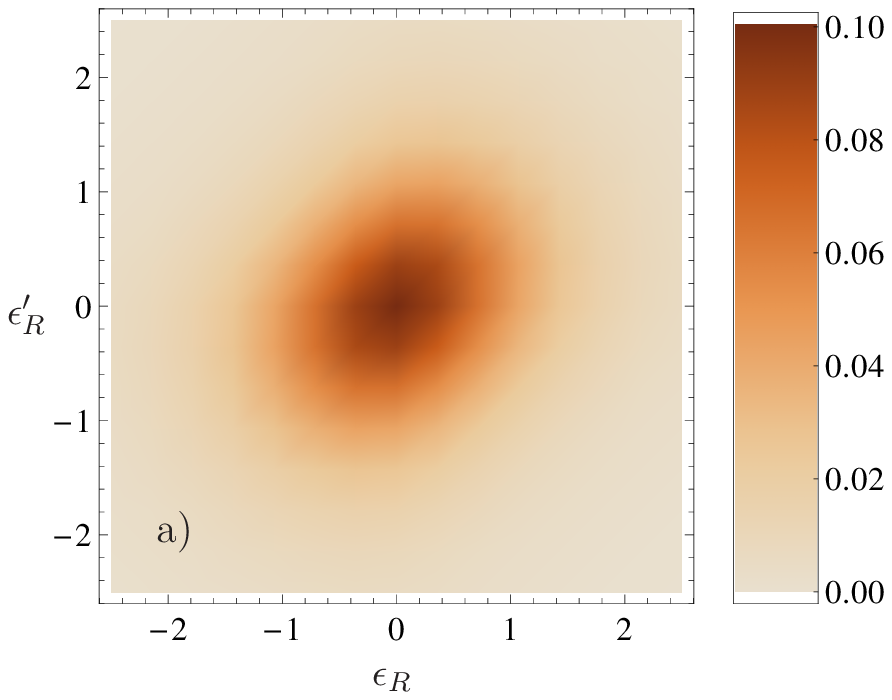}
\hspace{1cm}
\includegraphics[height=170pt]{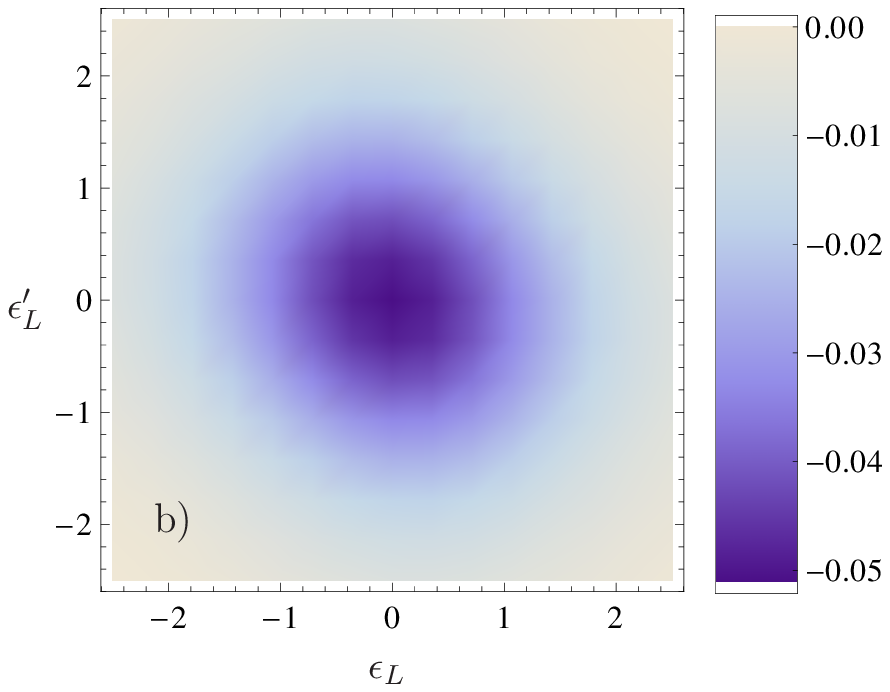}
\caption{\small (Color online). The energy dependence of the
  correlators $\langle\langle 
n_R(\epsilon_R, x_R)n_R(\epsilon_R', x_R')\rangle\rangle$ and  $\langle\langle
n_L(\epsilon_L, x_L)n_L(\epsilon_L', x_L')\rangle\rangle$ for
$x_\eta-x_{\eta}'=2Kl$. While the occupation numbers of the  hotter
(right in this case) electrons are correlated, the occupation numbers
of colder electrons are anti-correlated.  } 
\label{Max_cor_RR_LL}
\end{figure}

A similar analysis can be performed for the
case of the correlations in the occupation numbers of right electrons
along. One finds the correlation function of right distributions in
the form of 
(\ref{nRnR_form}) with
 \begin{multline}
f^{RR}_m(\epsilon_R, \epsilon_R', x)=
-\frac{V_F^2}{4}\int d\tau_R d\tau_R' 
	e^{ i\epsilon_L\tau_L+i\epsilon_R\tau_R }G^K_{R}(\tau_R)G^K_{R}(\tau_R')
	\left[\left(\frac{1-\frac{\ch\pi T_L\left( \tau_R+\tau_R'
                \right)}{\ch2\pi T_L x}} 
	{1-\frac{\ch\pi T_R\left( \tau_R+\tau_R' \right)}
	{\ch2\pi T_R x}}
	\frac{1 -\frac{\ch\pi T_R\left( \tau_R-\tau_R' \right)}{\ch2\pi T_R  x}}
	{1-\frac{\ch\pi T_L \left( \tau_R-\tau_R' \right)}{\ch2\pi T_L
            x}}\right)^{\gamma_m}-1 
	\right]\,.
\label{nRnR_ans}
\end{multline}
The exponents $\gamma_m$ are given by
\begin{equation}
 \gamma_m= {\cal T}^2 {\cal R}^{|m|}/(1+{\cal R}^2)\,.
\end{equation}
The analogous correlator of the left distributions can be obtained via
simple exchange $T_R\leftrightarrow T_L$. 
Provided that $T_R>T_L$ all the functions $f^{RR}_m$ turn out to be
positive, while the corresponding functions  
$f_m^{LL}$ are negative (cf. Fig. \ref{Max_cor_RR_LL} showing
$f_{2}^{RR}(\epsilon_R, \epsilon_R', 0)$ and  
$f_{2}^{LL}(\epsilon_L, \epsilon_L', 0)$). Thus, the hotter electrons
have positive correlations built into their distribution, while the
colder ones are  anti-correlated.

\section{Conclusions} 
\label{sec:conclusion}

In this paper we have developed an operator approach to the
non-equilibrium LL. Using bosonization and refermionization
techniques, we have explicitly determined the many-body density matrix of the
system and derived the fermionic correlation functions in
terms of Fredholm determinants $\Delta_{\eta}$.  Let us note that usually the correlation functions of the 
many-body interacting system can only be found approximately by truncating the Bogoliubov-Born-Green-Kirkwood-Yvon
chain. The model considered in this work constitutes a remarkable example of a many-body problem where all the correlation functions can be evaluated exactly. 

We have employed our technique to study the four-point correlation
functions of the electrons coming out of the LL wire. While the {\it
  dynamics} of the  outgoing electrons is free, the corresponding
density matrix is highly complicated. It incorporates the
correlations  
$\langle\langle n_{\eta}(x_\eta, \epsilon_\eta)n_{\eta'}(x_{\eta'},
\epsilon_{\eta'})\rangle\rangle$ in the electronic distribution
functions caused  by the scattering of the LL bosons at the boundaries
between the LL wire and the non-interacting leads.   
The spatial dependence of these correlations can be deduced from the
general analysis of the  corresponding functional determinants
$\Delta_{\eta}$. It  displays characteristic interference picks (or
dips)  of the width $l_{T^*}$ at distance $Kl$ (the ``optical
length'' of the interacting wire) one from another, see Fig.~\ref{RLX_dep}.   

For the case of partial equilibrium we have evaluated the Fredholm
determinants governing the correlation function 
$\langle\langle n_{\eta}(x_\eta, \epsilon_\eta)n_{\eta'}(x_{\eta'},
\epsilon_{\eta'})\rangle\rangle$ analytically. We have found
a non-trivial spatial dependence of correlations, both in the occupation
numbers of electrons of the same chirality and between the occupation numbers
of left- and right-movers.  Within the ``hotter'' chiral branch, the
positive correlation are developed in the electronic distribution. On
the other hand, anti-correlations are seen in the distribution of
colder electrons (see Fig. \ref{Max_cor_RR_LL}).  The sign of the
correlation function $\langle\langle n_{R}(x_R,
\epsilon_R)n_{L}(x_{L}, \epsilon_{L})\rangle\rangle$ depends on the
sign of $x_R+x_L$. On average,  
the occupation numbers of right- and left-movers are correlated at
energies of the same sign and anti-correlated at energies of opposite
signs, see Fig. \ref{Int_cor_RL}. The obtained results
indicate that quantum-interference effects contribute
crucially to the correlation functions. Intrinsically quantum correlations between parts of the system which can not be understood in a classical framework are referred to as entanglement\cite{Shroedinger}. In recent years it was recognized that quantifying 
the entanglement between subsystems  of a many-body quantum system can provide a clue to many relevant properties of the system, see e.g. Ref. \onlinecite{AmicoFazio,KlichLevitov}. Such quantification for a {\it mixed} state  of a quantum system is in general a highly complicated task \cite{Vedral}. It remains to be seen to what extent the correlations found in this work are relevant in the quantum information context.

We stress once again that the correlations studied are a genuine
non-equilibrium effect absent in equilibrium LL. These correlations 
are an
experimentally relevant quantity and can be measured 
in a specifically designed tunneling experiment. For example,
to access the correlator of left and right distribution functions, one
can imagine the following setup. Suppose that two tunneling probes
are attached to the wire at points $x_R$, $x_L$ satisfying $\eta
x_\eta>l/2$.  We can also assume for simplicity that the right probe
allows only the tunneling of right electrons,  while only left
electrons can tunnel through the left one. Then (under the assumption
that the tunneling densities states in the probes are not flat) the
correlator  
of right and left tunneling currents $\langle \langle
I_R(V_R)I_L(V_L)\rangle \rangle$ should be sensitive to the  
correlations in the distribution functions. The voltages $V_{\eta}$
biasing the probes will control the corresponding energies in
$\langle\langle n_{R}(x_R, \epsilon_R)n_{L}(x_{L},
\epsilon_{L})\rangle\rangle$.  
Clearly, a more sophisticated analysis is needed to extract the
information on the correlator of the distribution functions from the
correlation in the tunneling currents in a realistic setup. Our
analysis shows, however, that the correlator of the
distribution functions is in principle a measurable quantity.

Concluding the paper, we briefly discuss possible extentions
of the present work.  First, under the general non-equilibrium
conditions, a numerical analysis of 
the functional determinants $\Delta_{\eta}$ is needed to fully
understand  the correlations in the electronic  occupation
numbers. Second, the formalism developed here can be extended to cover
some non-stationary states of the LL.  
For example, in view of the recent experimental developments
\cite{Feve}, it is very interesting to investigate a LL wire exposed
to on-demand coherent single-electron sources.  In particular, in
the context quantum information processing, especially intriguing
is the entanglement generated by interaction between two electrons
injected by single-electron sources  from the left and right leads.  
Third, upon a proper modification, our technique should also be usefull for
the investigation of the fractional quantum Hall edge states out of
equilibrium. In this systems it is very interesting to look for
manifestation of the fractionally-charged quasiparticles in the
non-equilibrium correlation functions.

\section{Acknowledgements}

We thank P.M. Ostrovsky for useful discussions and acknowledge support by Alexander von Humboldt Foundation,
German-Israeli Foundation, DFG Center for Functional Nanostructures, and Israel Science Foundation.

\appendix

\section{Density matrix of non-equilibrium fermions in the bosonic representation}
\label{Density_matrix_appendix}

In this appendix, we transform the density matrix
(\ref{density_matrix_in_fermions}) into
bosonic representation. 
Throughout this Appendix we
will be dealing with the in-fermions $a^{in}$ and in-bosons $b^{in}$ only.
In order to simplify notations, we omit the index ``in'' for all
the operators. 
Further, we focus on right-moving in-fermions. The contribution of
right-movers is obtained analogously; the total density matrix
is the product of contributions of left and right in-particles.

We are thus looking for the bosonic representation of
the statistical operator 
\begin{equation}
	\hat{\rho}=\frac{1}{Z}\exp\left[ -\sum_{k}\epsilon(k)
          \left(a_{k}^+a_k-n^0(k)\right) \right]\,. 
\label{R}
\end{equation}
Here $Z$ is the normalization factor, $n^{0}(k)=\Theta(-k)$ is the
ground-state distribution function and  
$\epsilon(k)$ determines the distribution of the electrons via
\begin{equation}
 n(k)=\frac{1}{1+e^{\epsilon(k)}}\,.
\end{equation}

A particularly convenient basis in the Hilbert space of chiral
fermions  is provided by the  
bosonic coherent states which are the eigenstates of the
bosonic annihilation operators. Each coherent state is labeled by the
total number of fermions $N$  
and the set of eigenvalues $\beta_{q}$ of the operators $b_{q}$. Explicitly
\begin{equation}
 \left|N, \beta\right\rangle=
\exp\left[\sum_{q>0}\beta_{ q}b^+_{ q}\right]\left|N\right\rangle\,.
\label{coherent_states_def}
\end{equation}
The coherent states (\ref{coherent_states_def}) form the overcomplete
basis with the resolution of identity given by 
\begin{equation}
 1=\sum_{N}\int\left[ \prod_{q>0}d\beta^*_{q}d\beta_{q}\right]
\exp\left[-\sum_{q>0}\left|\beta_{q}\right|^2\right] 
\left|N, \beta\right\rangle
\left\langle N, \beta^*\right|\,.
\end{equation}
The overlap of two coherent states is given by
\begin{equation}
 \left\langle M, \beta^*\right|\left.N, \beta\right\rangle=\delta_{M,
   N}\exp\left[ 
\sum_{q>0}\left|\beta_q\right|^2
\right]\,.
\end{equation}

We are interested in the matrix elements of the statistical operator
$\hat{\rho}$ in the basis of the coherent states. They are given by
(obviously,  $\hat{\rho}$ is diagonal with respect to the total number
of fermions) 
\begin{equation}
	\langle \beta^*\,, N|\hat{\rho}|N\,,\beta\rangle=
	\langle N|\exp\left[\sum_{q>0}\beta_q^* b_q\right]\hat{\rho}
	\exp\left[ \sum_{q>0}\beta_q b^+_q \right]|N\rangle\,.
\label{mat_elem_basic}
\end{equation}
We remind that the bosons are proportional to the Fourier components
of the fermionic density:
\begin{eqnarray}
b^+_{ q}=
\sqrt{\frac{2\pi}{L |q|}}\sum_{k}a^+_{k+q}a_{ k}\,, \qquad
b_{ q}=\sqrt{\frac{2\pi}{L |q|}}\sum_{k}a^+_{k-q}a_{ k}\,.
\end{eqnarray}
Thus, the operator under the average in (\ref{mat_elem_basic}) is
exponentially quadratic in fermions.  

It is convenient to introduce new fermions:
\begin{eqnarray}
	c_k=a_k \left( 1-n^{N}(k) \right)+a_k^+n^N(k)\,,\\
	c_k^+=a_k^+\left( 1-n^N(k)\right)+a_kn^N(k)\,.
\label{trans}
\end{eqnarray}
Here $n^{N}(k)=\Theta\left( -k+\frac{2\pi}{L}N \right)$ is the
distribution function of a Fermi sea with $N$ extra particles. The idea
behind the transformation (\ref{trans}) is that the state $|N\rangle $
is nullified  by all the operators $c_k$, which simplifies the
derivation. In terms of the new fermions $c$ we have  
\begin{equation}
	\sum_{q>0}\beta_{q}b_q^+=\sum_{k_1\,, k_2} 
	c^+_{k_1}c_{k_2}U_{k_1, k_2}+\sum_{k_1\,,
          k_2}c_{k_1}^+c_{k_2}^+ V_{k_1, k_2}\,. 
\end{equation}
The matrices $U_{k_1, k_2}$ and $V_{k_1,k_2}$ carring two momentum
indices are given by 
\begin{eqnarray}
	U=\Phi (1-n^N)-\Phi^T n^N\,,\\
	V=\frac12\left[ \Phi+\Phi^T, n^N\right]\,.
\end{eqnarray}
Here the distribution function $n^N$ is considered as a matrix
diagonal in momentum space, while  
the matrix $\Phi$ has the matrix elements
\begin{equation}
 \Phi_{k_1-k_2}= \Theta(k_1-k_2)\sqrt{\frac{2\pi}{L (k_1-k_2)}}\beta_{k_1-k_2}\,. 
\end{equation}
Using the fermionic commutation relations satisfied by $c_k$ and the
fact that all $c_k$ annihilate $|N\rangle$, one  now finds 
\begin{eqnarray}
	\exp\left[ \sum_{q>0}\beta_q b^+_q \right]|N\rangle=
	\exp\left[ 
	\sum_{k_1k_2}c_{k_1}^+c_{k_2}^+ W_{k_1k_2}\right]|N\rangle\,,\\
	W=\int_0^1 ds \exp\left[ sU \right]V\exp\left[s U^T \right]\,.
\end{eqnarray}
Note that the operator $\Phi$ preserves the subspace 
$k>\frac{2\pi}{L}N$ while the operator $\Phi^T$ preserves the subspace
$k\leq\frac{2\pi}{L}N$.  
On the other hand, on the first of these subspaces 
$1-n^N=1$, while the other subspace is nullified by $1-n^N$. 
It follows that 
\begin{equation}
	e^{s U}=e^{s\Phi}(1-n^N)+e^{-s \Phi^T} n^N\,.
\end{equation}
Using this relation, the expression for the matrix $W$ can be
simplified to the form
\begin{equation}
	W=
	\frac12 \left[ 
	e^{-\Phi^T}\left( 1-n^N \right)e^{\Phi^T}-e^{\Phi}\left( 1-n^N \right)
	e^{-\Phi}
	\right]\,.
\end{equation}

The matrix element (\ref{mat_elem_basic}) now reads
\begin{multline}
	\langle \beta^*\,,
        N|\hat{\rho}|N\,,\beta\rangle=\frac{\gamma}{Z}\left \langle
          N\right|\exp\left[  
	\sum_{k_1k_2}c_{k_1}c_{k_2} W^+_{k_1k_2}\right] 
	\exp\left[  -\sum_{k_1k_2}
	c_{k_1}^+c_{k_2}E_{k_1k_2} \right]
	\exp\left[ 
	\sum_{k_1k_2}c_{k_1}^+c_{k_2}^+ W_{k_1k_2}\right]\left| N\right\rangle\\
	\equiv \frac{\gamma}{Z}\left \langle
          N\right|e^{\widehat{W}^+}e^{\widehat{E}} 
	e^{\widehat{W}}\left| N\right\rangle\,,
\label{c_ans}
\end{multline}
where
\begin{equation}
	E_{k_1, k_2}=\epsilon(k_1)\left( 1-2n^N(k_1)
        \right)\delta_{k_1k_2}\,,\qquad 
	\gamma=\exp\left[ -\sum_k\epsilon(k)\left( n^{N}(k)-n^0(k)
          \right) \right]\,. 
\end{equation}  
For the evaluation of  the average  (\ref{c_ans}) one can employ the
standard technique of the fermionic coherent states. Let us define the
set of the eigenstates of annihilation operators $c_k$ with the
Grassmann eigenvalues  
$\chi_k$ ($\chi$ stands for the full set of $\chi_k$ with
$k=-\infty\ldots \infty$): 
\begin{equation}
 c_k|\chi\rangle=\chi_k|\chi\rangle\,.
\end{equation}
The coherent states $|\chi\rangle$  form an overcomplete basis with
the resolution of identity 
\begin{equation}
 1=\int D\chi^* D\chi
 \exp\left[-\sum_{k}\chi^*_k\chi_k\right]|\chi\rangle\langle\chi^*|\,, 
\label{identity}
\end{equation}
where 
\begin{equation}
 D\chi^* D\chi =\prod_{k}d\chi^*_kd\chi_k\,.
\end{equation}
Using the resolution of identity (\ref{identity}) and noting that the
state $|N\rangle$ is itself a coherent state  
$|\chi=0\rangle$, we get
\begin{multline}
 	\langle \beta^*\,, N|\hat{\rho}|N\,,\beta\rangle=\frac{\gamma}{Z}
	\int D\chi^{1*}D\chi^{1}
        D\chi^{2*}D\chi^{2}\exp\left[-\sum_{k}\left(\chi^{1*}_k\chi^1_k+\chi^{2*}_k\chi^2_k\right)\right]    
	\\\times\left \langle
          \chi^*=0\right|e^{\widehat{W}^+}|\chi^1\rangle\langle
        \chi^{1*}|e^{\widehat{E}} 
	|\chi^2\rangle\langle \chi^{2*}|
	e^{\widehat{W}}\left| \chi=0\right\rangle\,.
\end{multline}
Finally, since
\begin{eqnarray}
 \langle\chi^* |e^{\widehat{W}}|\chi\rangle=\exp\left[
\sum_{k}\chi_k^*\chi_k+\sum_{k_1, k_2}\chi_{k_1}^*\chi_{k_2}^*W_{k_1, k_2}\right]\,,\\
 \langle\chi^* |e^{\widehat{E}}|\chi\rangle=\exp\left[
\sum_{k_1, k_2}\chi_{k_1}^*\chi_{k_2}^*\left[e^{-E}\right]_{k_1k_2}\right]\,,\\
\end{eqnarray}
we are left with a simple Gaussian integral over Grassman variables.
The result of the integration reads 
\begin{equation}
	\langle \beta^*, N|\hat{\rho}|N,\beta\rangle=\frac{\gamma}{Z} \left[ 
	\det
	\left(
	\begin{array}{cc}
		1 & -2 e^{-E}We^{-E}\\
		2 W^+ & 1
	\end{array}
	\right)
	\right]^{1/2}\,.
\end{equation}
Using the explicit form of the matrices $W$ and $E$, one can reduce the
expression above to  
\begin{equation}
	\langle \beta^*, N|\hat{\rho}|N,\beta\rangle
	=\frac{\gamma}{Z}
	\det \left[ 
	e^{-\epsilon}e^{\Phi}\left(1-n_N\right) 
	e^{-\Phi}e^{\epsilon}+e^{-\Phi^+}n_Ne^{\Phi^+}
	\right]\,.
\label{R_final_ans}
\end{equation}
Finally, working out the factor in front of the determinant in
(\ref{R_final_ans}), we find 
\begin{equation}
	\langle \beta^*, N|\hat{\rho}|N, \beta\rangle
	=\det\left[1- 
	e^{\Phi}n_N 
	e^{-\Phi}(1-n)-
	e^{-\Phi^+}(1-n_N)e^{\Phi^+}n
	\right]\,.
\label{R_final_ans_final}
\end{equation} 
Equation (\ref{R_final_ans_final}) gives an explicit expression for
the density matrix of non-equilibrium interacting fermions in the bosonic
basis. It has a form of a one-dimensional functional
determinant of Fredholm type.

Let us verify that  Eq.~(\ref{R_final_ans_final}) reduces to the
Boltzmann-Gibbs form at equilibrium, which
in present notations corresponds to $\epsilon(k)=kV_F/T$.  
Indeed, we know that for $\epsilon(k)\equiv 0$ Eq. (\ref{R_final_ans})
should give just the norm of the coherent  
$|N, \beta\rangle$ (up to the normalization factor $Z$)
\begin{equation}
 	\left.\langle \beta^*, N|\hat{\rho}|N,\beta\rangle\right|_{\epsilon(k)=0}
	=\frac{\gamma}{Z}
	\det \left[ 
	e^{\Phi}\left(1-n_N\right) 
	e^{-\Phi}+e^{-\Phi^+}n_Ne^{\Phi^+}
	\right]=\frac{1}{Z}\exp\left[\sum_{q>0}\left|\beta_q\right|^2\right]\,.
\end{equation}
On the other hand for $\epsilon(k)=kV_F/T$ we can write
\begin{equation}
	\langle \beta^*, N|\hat{\rho}_{eq}|N,\beta\rangle
	=\frac{\gamma}{Z}
	\det \left[ 
	e^{-\epsilon}e^{\Phi}e^{\epsilon}\left(1-n_N\right) 
	e^{-\epsilon}e^{-\Phi}e^{\epsilon}+e^{-\Phi^+}n_Ne^{\Phi^+}
	\right]=
	\frac{\gamma}{Z}
	\det \left[ 
	e^{\widetilde{\Phi}}\left(1-n_N\right) 
	e^{-\widetilde{\Phi}}+e^{-\Phi^+}n_Ne^{\Phi^+}
	\right]\,,
\end{equation}
where
\begin{equation}
 \widetilde{\Phi}_{k_1-k_2}=\left[e^{-\epsilon}\Phi e^{\epsilon}\right]_{k_1, k_2}=
\Theta(k_1-k_2)\sqrt{\frac{2\pi}{L (k_1-k_2)}}\beta_{k_1-k_2}e^{-V_F(k_1-k_2)/T}\,.
\end{equation}
We see that the matrix elements of the equilibrium density matrix can
be obtained from the norm of the coherent states just by the
replacement 
\begin{equation}
 \beta_q\rightarrow \beta_q e^{-V_F q/T}\,,
\end{equation}
with the result
\begin{equation}
	\langle \beta^*, N|\hat{\rho}_{eq}|N,\beta\rangle=
	\frac{1}{Z}\exp\left[\sum_{q>0}e^{-q
            V_F/T}\left|\beta_q\right|^2\right]\,. 
	\label{R_eq_final}
\end{equation}
 In the operator form, Eq. (\ref{R_eq_final}) reads
\begin{equation}
 \hat{\rho}_{eq}=\frac{1}{Z}\exp\left[-\frac{V_F}{T}\sum_{q>0}q
   b^+_{q}b_{q}\right] \,,
\end{equation}
which is the expected equilibrium result.

Away from equilibrium, the bosonic density
matrix (\ref{R_final_ans_final}) is much more complicated. Nevertheless,
it can be used to derive the expressions
(\ref{prototypical_function_answer},  \ref{Delta_Def},
\ref{delta_def})  for the prototypical average
(\ref{prototypical_function}) required for evaluation of correlation
functions.   
To do this, one represents the determinant (\ref{R_final_ans_final}) as
an a Gaussian integral over auxiliary Grassmann variables and performs
an expansion in powers of $e^{\Phi}$. At every order,  evaluation of the
correlator of the bosonic exponents (\ref{prototypical_function}) is
then given by a Gaussian integral over  
bosons and fermions. The crucial point is some specific cancellations
between the fermionic and bosonic integrations, which finally allows us
to evaluate all orders of the expansion analytically and resum the
whole series.   This brute-force  derivation requires, however, 
rather cumbersome combinatorics,  and we do not present it here. 
An alternative, considerably simpler, derivation 
involves refermionization, as explained in the main text
and Appendix \ref{trace}.

\section{Averaging exponential of bosonic fields in a
  non-equilibrium Luttinger liquid via refermionization}
\label{trace}

In this Appendix we evaluate the prototypical correlation function
(\ref{prototypical_function}) where the average is performed with the
density matrix (\ref{density_matrix_in_fermions}). Since the
right-moving in-fermions are completely independent from the
left-moving ones (see Eq. (\ref{density_matrix_in_fermions})) we
ignore the latter 
for a while and include the left-movers in the final formulas. In our
notations we also suppress  
the index ''in'' since all the operators (bosons and fermions) we will
be dealing with in this Appendix  are in-operators and the suppression
should not cause any confusion. 
Thus, we need to evaluate the average
\begin{equation}
Z[g^*, g]=\left\langle \exp\left[\sum_{q>0}g(q)b^{+}_{q} \right] 
	\exp\left[-\sum_{ q>0}g^*(q)b_{q} \right]\right\rangle
\label{av}
\end{equation}
with the density matrix
\begin{equation}
 \hat{\rho}=\frac{1}{Z}\exp\left[-\sum_{k}\epsilon(k)
	\left(a^{+}_{ k}a_{ k}-n^0(k)\right)\right]\,.
\label{R_appendix}
\end{equation}
We can now apply the well-known expression for  the trace over the
fermionic Hilbert space of a product of exponentially quadratic  
operators\cite{Klich}:  
 \begin{equation}
  \tr e^{H_1}\ldots e^{H_n}=\det\left( 1 +e^{h_1}\ldots e^{h_n}\right)\,,
\label{trace-det}
 \end{equation}
where
\begin{equation}
 H_i=\sum_{k_1, k_2} h_i^{k_1, k_2}a^+_{k_1}a_{k_2}\,.
\end{equation}
The right-hand side of (\ref{trace-det}) is a determinant of an
operator acting in the single-particle Hilbert space. 
Applying (\ref{trace-det}) to the average in question, we  get
\begin{eqnarray}
Z[g^*, g]=\det\left( 1-n(k)+e^{-i\delta(x)}n(k) \right)\,,
\label{Z_naiv}
\\
\delta(x)=i\sum_{q>0}\sqrt{\frac{2\pi}{L q}}\left(
g_{ q} e^{iqx}-g^*_{q}e^{-iqx}
\right)	\,.
\end{eqnarray}

The determinant in (\ref{Z_naiv}) is in fact not well defined, for the
following reason. Strictly
speaking, Eq.~(\ref{trace-det}) assumes that in the fermionic
Hilbert space there is a state nullified by all the operators
$a_k$. This is not the case in the present situation. 
We can resolve this difficulty 
by making the transformation from the particle operators $a_k$,
$a_k^+$ to particle and hole operators: 
\begin{eqnarray}
	c_k=a_k \left( 1-n^{0}(k) \right)+a_k^+n^0(k)\,,\\
	c_k^+=a_k^+\left( 1-n^0(k)\right)+a_kn^0(k)\,.
\end{eqnarray}
Then we can evaluate $Z[g^*, g]$ with the technique of fermionic
coherent states. The calculation is very similar to that of Appendix
\ref{Density_matrix_appendix}.  

We can make a short-cut, however, if we notice that $Z[g^*, g]$ is an
average of an operator normal-ordered in bosons. 
Thus, $Z[g^*, g]=1$ at zero temperature. This  suggests  that the
regularized version of Eq.~(\ref{Z_naiv}) should be 
\begin{equation}
 Z[g^*, g]=\det\left[\left( 1-n^0(k)+e^{-i\delta(x)}n^0(k)
   \right)^{-1}\left( 1-n(k)+e^{-i\delta(x)}n(k) \right)\right] 
\label{Z_true}
\end{equation}
A direct calculation along the lines of Appendix
\ref{Density_matrix_appendix} indeed confirms this result. 

Incorporating now the left electrons into our consideration, we
come to Eqs. (\ref{prototypical_function_answer},
\ref{Delta_Def}, \ref{delta_def}).


\begin{thebibliography}{99}
\bibitem{Luttinger}
J. M. Luttinger, J.Math. Phys. {\bf 4}, 1154 (1963).
\bibitem{Bockrath}
M. Bockrath, D. H. Cobden, J. Lu, A. G. Rinzler, R. E.  Smalley, L.  Balents, and P. L. McEuen, Nature (London) {\bf 397}, 598 (1999).
\bibitem{Yao}
Z. Yao,  H. W. Ch. Postma,  L.  Balents, , and C.  Dekker, Nature {\bf 402}, 273 (1999).
\bibitem{Schoenenberger}
C. Sch{\"o}nenberger, Semicond. Sci. Technol {\bf 21}, S1 (2006).
\bibitem{Auslaender}
O. M. Auslaender, A.  Yacoby, R. de Picciotto, K. W. Baldwin, L. N.   Pfeifer,  and K. W.  West, Science {\bf 295}, 825 (2002).
\bibitem{Slot}
  E. Slot, M. A. Holst, H. S. J. van der Zant,  and S. V. Zaitsev-Zotov, Phys. Rev. Lett. {\bf 93}, 176602 (2004).
\bibitem{Picciotto} 
R. de Picciotto, M. Reznikov, M. Heiblum, V. Umansky, G. Bunin, and D. Mahalu, Nature {\bf 389}, 162 (1997).
\bibitem{Grayson}
  M. Grayson, L. Steinke, D. Schuh, D.  M.  Bichler, L. Hoeppel, J. Smet, K. v. Klitzing, D. K. Maude, and G. Abstreiter, Phys. Rev. B {\bf 76}, 201304(R)  (2007).
\bibitem{Chang}
  A. M. Chang, Rev. Mod. Phys. {\bf 75}, 1449 (2003).
\bibitem{GiamarchiTsvelik}
T. Giamarchi and A. M. Tsvelik,  Phys. Rev. B {\bf 59}, 11398 (1999).
\bibitem{Chitra}
R. Chitra and T. Giamarchi, Phys. Rev. B {\bf 55}, 5816 (1997)
\bibitem{Luther}
 A. Luther and  I. Peschel, Phys. Rev. B {\bf 9}, 2911 (1974).
\bibitem{Coleman}
  S. Coleman, Phys. Rev. D {\bf 11}, 2088, (1975).
\bibitem{Mandelstam} 
  S. Mandelstam, Phys. Rev. D {\bf 11}, 3026 (1975).
\bibitem{MattisLieb} 
D.C. Mattis and E.H. Lieb, J. Math. Phys. {\bf 6}, 304 (1965).
\bibitem{Mattis} 
D.C. Mattis, J. Math. Phys. Phys. {\bf 15}, 609 (1974).
\bibitem{Heidenreich} 
R. Heidereich, B. Schroer, R. Seiler, and D. Uhlenbrock, Phys. Lett. A {\bf 54}, 119 (1975).
\bibitem{Haldane} 
F. D. M. Haldane, J. Phys. C {\bf 14}, 2585 (1981).
\bibitem{Delft} 
J. von Delft and H. Schoeler, Annalen Phys. {\bf 7}, 225 (1998).

\bibitem{ChuiLee} 
  S. -T. Chui and P. A. Lee, Phys. Rev. Lett. {\bf 35}, 315 (1975).
\bibitem{Voit} 
J. Voit, Rep. Prog. Phys. {\bf 58}, 977 (1995).
\bibitem{KaneFisher}
  C. L. Kane and M. P. A. Fisher, Phys. Rev. B {\bf 46}, 15233 (1992).
\bibitem{Voit1}
 J. Voit, Phys. Rev. B {\bf 45}, 4027 (1992).

\bibitem{Giamarchi}
T. Giamarchi, {\it Quantum Physics in One Dimension}, (Clarendon Press,  Oxford, 2004).
\bibitem{ChenDirks}
Y.-Fu Chen,   T. Dirks,   G. Al-Zoubi,   N. O. Birge,  and N. Mason, Phys. Rev. Lett. {\bf 102}, 036804 (2009).
\bibitem{Dirks}
T. Dirks, Y.-F.  Chen, N.-O.  Birge, and N.  Mason, Appl. Phys. Lett. {\bf 95}, 192103 (2009).
\bibitem{Altimiras}
C. Altimiras,  H. le Sueur, U. Gennser, A. Cavanna,  D. Mailly, and F. Pierre, Phys. Rev. Lett. {\bf 105}, 226804 (2010).
\bibitem{LeSueurAltimiras}
H. le Sueur, C. Altimiras, U. Gennser, A. Cavanna, D. Mailly, and F. and Pierre, Phys. Rev. Lett. {\bf 105}, 056803 (2010).
\bibitem{LeSueur} 
C. Altimiras, H. le Sueur, U. Gennser, A. Cavanna, D. Mailly and F. Pierre, Nature Physics {\bf 6}, 34 (2010).


\bibitem{Gutman1}
D. B. Gutman, Y. Gefen,  and A. D. Mirlin, Phys. Rev. Lett. {\bf 101}, 126802 (2008); Phys. Rev. B {\bf 80}, 045106 (2009).
\bibitem{Gutman3}
D. B. Gutman, Y. Gefen,  and A. D. Mirlin, Europhys. Lett. {\bf 90}, 37003 (2010); Phys. Rev. B {\bf 81}, 085436 (2010).
\bibitem{Gutman4}
D. B. Gutman, Y. Gefen,  and A. D. Mirlin, Phys. Rev. Lett. {\bf 105}, 256802 (2010).
\bibitem{Gutman11}
D. B. Gutman,   Y. Gefen, and A. D. Mirlin, J. Phys. A: Math. Theor. {\bf 44}, 165003 (2011).
\bibitem{Jacobs07} 
S. G. Jakobs,  V. Meden, and H. Schoeller, Phys. Rev. Lett. {\bf 99}, 150603 (2007).
\bibitem{NgoDinh10}
S. Ngo~Dinh, D. A. Bagrets, and A.D. Mirlin, Phys. Rev. B {\bf 81}, 081306(R) (2010).
\bibitem{Takei10}
S. Takei, M. Milletar, and B. Rosenow, Phys. Rev. B {\bf 82}, 041306(R) (2010).
\bibitem{Bena10}
C. Bena, Phys. Rev. B {\bf 82}, 035312 (2010).
\bibitem{Pugnetti09}
S. Pugnetti,  F. Dolcini, D. Bercioux, and H. Grabert, Phys. Rev. B {\bf 79}, 035121 (2009).
\bibitem{Trushin08}
M. Trushin and A. L. Chudnovskiy, Europhys. Letters {\bf 82}, 17008 (2008).



\bibitem{Chalker07}
J. T. Chalker,  Y. Gefen, and M.Y. Veillette, Phys. Rev. B {\bf 76}, 085320 (2007).
\bibitem{Levkivskiy08}
I. P. Levkivskyi and E. V. Sukhorukov, Phys. Rev. B {\bf 78}, 045322 (2008);
Phys. Rev. Lett. {\bf 103}, 036801 (2009).
\bibitem{Kovrizhin10} D. L. Kovrizhin and J. T. Chalker, Phys. Rev. B {\bf 81}, 155318 (2010); 
arXiv:1009.4555 (2010). 
\bibitem{Schneider11}
M. Schneider,  D. A. Bagrets, and A. D. Mirlin, arXiv:1101.4391;  to appear in Phys. Rev. B  (2011).


\bibitem{Levitov1} L. S. Levitov and G. B. Lesovik, JETP Lett. {\bf 58}, 230 (1993);
L. S. Levitov, H. Lee, and G. B. Lesovik, J. of Math. Phys. {\bf 37}, 4845 (1996).
\bibitem{Abanin04}
D. A. Abanin and L. S. Levitov, Phys. Rev. Lett. {\bf 93}, 126802 (2004); Phys. Rev. Lett. {\bf 94}, 186803 (2005).
\bibitem{Snyman07}
I. Snyman and Y. V. Nazarov, Phys. Rev. Let. {\bf 99}, 096802 (2007).



\bibitem{Maslov}
D. L. Maslov  and M. Stone, Phys. Rev. B {\bf 52}, 5539(R) (1995).
\bibitem{Ponomarenko}
V. V. Ponomarenko, Phys. Rev. B {\bf 52}, 8666(R) (1995); {\bf 54} 10328, (1996).
\bibitem{Trauzettel}
B. Trauzettel,  I. Safi,  Fa. Dolcini,   and   H. Grabert Phys. Rev. Lett. {\bf 92}, 226405 (2004).
\bibitem{Glazman} 
G. Barak, H. Steinberg, L. N.  Pfeiffer, K. W.   West, L. Glazman, F. von Oppen, and A. Yacoby, Nature Physics {\bf 6}, 489 (2010).
\bibitem{comment_arguments}
There is no need to consider more general arguments of the $\psi$ operators because
 in the non-interacting leads each $\psi_{\eta}(x, t)$ is the function of $x-\eta V_F t$ only.
\bibitem{comment2}
 In our consideration of partial equilibrium we set the chemical potentials $\mu_{R(L)}$ of the left- and right-movers equal to zero. Non-zero $\mu_\eta$ would lead to trivial energy shifts in our formulas. 
\bibitem{Shroedinger} E. Schr{\"o}dinger, Proc. Cambridge. Phil. Soc. {\bf 31}, 555 (1935).
\bibitem{AmicoFazio}
 L. Amico, R. Fazio, A. Osterloh, and  V. Vedral,  Rev. Mod. Phys. {\bf 80}, 517 (2008).
\bibitem{KlichLevitov} 
I. Klich and L.S. Levitov, Phys. Rev. Lett. {\bf 102}, 100502 (2009).
\bibitem{Vedral} For the review of the entanglement measures see V. Vedral, Rev. Mod. Phys. {\bf 74}, 197 (2002).
\bibitem{Feve}
G. F{\`e}ve, A. Mah{\'e}, J.-M.  Berroir, T.  Kontos, B.  Pla\c{c}ais, D. C. Glattli, A. Cavanna, B. Etienne, and Y.  Jin, Science {\bf 316}, 1169 (2007).
\bibitem{Klich} I. Klich, in {\it Quantum Noise in Mesoscopic Systems}, ed. by Yu. V. Nazarov (Kluwer, Dordrecht, 2003); cond-mat/0209642.


\end{thebibliography}
\end{document}